\documentclass[fleqn]{aa}  

\usepackage{graphicx}
\usepackage{caption}
\usepackage{subcaption}
\usepackage{booktabs}
\usepackage{pgfplotstable}
\usepackage{colortbl}
\usepackage{xcolor}
\usepackage{soul} 
\usepackage{amsmath}
\pgfplotstableset{
    every head row/.style={
    before row=\toprule,after row=\midrule},
    every last row/.style={
    after row=\bottomrule},
    col sep = &,
    row sep=\\,
    string type,
}
\usepackage{txfonts}

\usepackage{natbib,twoopt}
\usepackage[breaklinks=true]{hyperref}
\bibpunct{(}{)}{;}{a}{}{,}
\makeatletter
  \newcommandtwoopt{\citeads}[3][][]{\href{http://adsabs.harvard.edu/abs/#3}%
    {\def\hyper@linkstart##1##2{}%
     \let\hyper@linkend\@empty\citealp[#1][#2]{#3}}}
  \newcommandtwoopt{\citepads}[3][][]{\href{http://adsabs.harvard.edu/abs/#3}%
    {\def\hyper@linkstart##1##2{}%
     \let\hyper@linkend\@empty\citep[#1][#2]{#3}}}
  \newcommandtwoopt{\citetads}[3][][]{\href{http://adsabs.harvard.edu/abs/#3}%
    {\def\hyper@linkstart##1##2{}%
     \let\hyper@linkend\@empty\citet[#1][#2]{#3}}}
  \newcommandtwoopt{\citeyearads}[3][][]%
    {\href{http://adsabs.harvard.edu/abs/#3}
    {\def\hyper@linkstart##1##2{}%
     \let\hyper@linkend\@empty\citeyear[#1][#2]{#3}}}
\makeatother
\begin{document}

   \title{Externally irradiated young stars in NGC 3603}

   \subtitle{A JWST NIRSpec catalogue of pre-main-sequence stars in a massive star formation region}

   \author{Ciarán Rogers
          \inst{1},
          Guido de Marchi,
          \inst{2},
          Bernhard Brandl
          \inst{1, 3}}

   \institute{Leiden Observatory, Leiden University,
              PO Box 9513, 2300 RA Leiden, The Netherlands\\
              \email{rogers@strw.leidenuniv.nl}
         \and
             European Space Research and Technology Centre, Keplerlaan 1, 2200 AG Noordwijk, The Netherlands\\
             \email{gdemarchi@rssd.esa.int}
         \and
            Faculty of Aerospace Engineering, Delft University of Technology, Kluyverweg 1, 2629 HS Delft, The Netherlands\\
            \email{brandl@strw.leidenuniv.nl}}

   \date{Received December 9, 2024; accepted April 7, 2025}
 
  \abstract
   {NGC 3603 is the optically brightest massive star forming region (SFR) in the Milky Way, representing a small scale starburst region. Studying young stars in regions like this allows us to assess how star and planet formation proceeds in a dense clustered environment with high levels of UV radiation. JWST provides the sensitivity, unbroken wavelength coverage, and spatial resolution required to study individual pre-main-sequence (PMS) stars in distant massive SFRs in detail for the first time.}
   {We identify a population of accreting PMS sources in NGC 3603 based on the presence of hydrogen emission lines in their NIR spectra. We spectrally classify the sources, and determine their mass and age from stellar isochrones and evolutionary tracks. From this we determine the mass accretion rate $\dot{M}_{acc}$ of the sources and compare to samples of stars in nearby low-mass SFRs. We search for trends between $\dot{M}_{acc}$ and the external environment.}
   {Using the micro-shutter assembly (MSA) on board NIRSpec, multi-object spectroscopy was performed, yielding 100 stellar spectra. Focusing on the PMS spectra, we highlight and compare the key features that trace the stellar photosphere, protoplanetary disk, and accretion. We fit the PMS spectra to derive their photospheric properties, extinction, and NIR veiling. From this, we determined the masses and ages of our sources by placing them on the Hertzsprung-Russel diagram (HRD). Their accretion rates were determined by converting the luminosity of their hydrogen emission lines to an accretion luminosity.}
   {Of the 100 stellar spectra obtained, we have classified 42 as PMS and actively accreting. Our sources span a range of masses from 0.5 to 7 $M_{\odot}$. Twelve of these accreting sources have ages consistent with $\ge$ 10 Myr, with four having ages of $\ge$ 15 Myr. The mass accretion rates of our sample span 5 orders of magnitude and are systematically higher for a given stellar mass than for a comparative sample taken from low-mass SFRs. We report a relationship between $\dot{M}_{acc}$ and the density of interstellar molecular gas as traced by nebular $H_2$ emission.}
   {}


   \maketitle

\section{Introduction}
\label{sec:intro}
The majority of stars in the Milky Way formed in clusters that contained at least one massive star ($M_{*} \ge 8 M_{\odot}$) \citep{lada2003embedded}. It is likely that our own solar system formed in such an environment \citep{pfalzner2015formation}. Fundamental differences exist between the environments of massive and low-mass SFRs, in particular with respect to the intensity of the UV radiation field and the interstellar gas and dust density.\\ 
The majority of spectroscopy-based studies of PMS stars have looked towards low-mass SFRs, in part motivated by their proximity. These regions tend to be within a few hundred parsecs (pc), making it feasible to study the PMS stars with high signal to noise (S/N), and to spatially resolve features like outflows and protoplanetary disks. However, despite the advantages offered by these nearby regions, they do not represent the typical star and planet formation environment found in the Milky Way, or throughout the Universe. The lack of high mass stars, low stellar densities, and modest interstellar gas and dust densities are in sharp contrast to clustered star forming environments. The closest example of clustered star formation is in Orion, located at a distance of $414$ pc \citep{2007A&A...474..515M}. Orion, which contains several sights of massive star formation, most notably the Orion Nebula Cluster (ONC), is the most well studied massive SFR in the Milky Way \citep[e.g.][]{ricci2008hubble, megeath2012spitzer, lewis2016protostars, berne2023formation}. Even Orion represents a relatively modest cluster, with the ONC containing only a handful of stars with spectral type O, the most massive of which is $\theta^1\;Ori\;C$, an O6 star lying at the heart of the Trapezium cluster. In contrast, NGC 3603 contains upwards of $40$ O-type stars, including several Wolf-Rayet stars, all located within an extremely compact volume in the centre of the cluster \citep{drissen1995dense}. These massive stars inject tremendous radiative and mechanical feedback into their environment, ionising their surroundings to an extent that is several orders of magnitude more extreme compared to Orion, which is itself several orders of magnitude more extreme than low-mass SFRs. Like the ONC, there remain dense photodissociation regions (PDRs) consisting of ionised, atomic and molecular material surrounding the stars in  NGC 3603. The impact that these prominent environmental factors have on star and planet formation remains largely unknown, though recently more attention has been given to the potential implications.\\ 
In \cite{o1993discovery}, the authors reported on the discovery of a population of externally irradiated protoplanetary disks in the ONC, dubbed proplyds. The UV radiation from the massive stars was seen to be driving powerful thermal winds from the surface of nearby protoplanetary disks, with mass loss estimates of $10^{-6}\;M_{\odot} yr^{-1}$. These mass loss rates would lead to disk dispersal in $\sim 10^5$ years, dramatically shorter than the dispersal timescales seen in low-mass SFRs $\sim 5$Myr \citep{mamajek2009initial}. In \cite{haworth2018fried, haworth2023fried}, the external photoevaporative mass loss rates of protoplanetary disks were calculated for a range of stellar masses, disk radii, and external UV field strengths. The UV field strength is given in Habing units $G_{0}$, with the typical interstellar medium having a value of 1 $G_{0}$. The authors found that in $G_0$ environments of $\sim 10^4$, comparable to the ONC and lower than in NGC 3603, disks around low and intermediate mass stars exhibit high mass loss rates down to disk radii of just $1$AU. In \cite{walsh2012chemical}, simulations were performed on an externally irradiated protoplanetary disk to understand the chemical changes that take place in the disk due to UV and X-ray radiation. They found that the UV radiation leads to a depletion of molecules in the inner disk ($\le1$AU), and an enhancement of molecules outside of this radius. In \cite{kuffmeier2023rejuvenating} the authors simulated star formation in regions of high gas and dust density, with column densities of interstellar material comparable to that found in NGC 3603 \citep{nurnberger2002molecular, schneider2015understanding, rocamora2024exploring}. The aim of this study was to investigate whether late-stage infall of interstellar material could rejuvenate protoplanetary disks, and possibly affect the evolution of the central star itself. They found that in such dense clustered environments not only does late stage infall commonly occur, but that in these episodes, on average $\ge0.1\;M_{\odot}$ is added to the star-disk system. For stars with a final mass of $\ge 1\;M_{\odot}$, more than $50\;\%$ of their mass comes from late stage infall. In \cite{winter2024spatially}, the authors reported a spatial correlation between the mass accretion rate and interstellar molecular gas density for a sample of young stars in Lupus, appearing to be in agreement with the results from \cite{kuffmeier2023rejuvenating} that mass accretion can be influenced by the external environment. \\
This non-exhaustive list of studies highlights the need to consider the role of the external environment in the context of star and planet formation. Not only because there is compelling evidence that it can be a dominant influence on the evolution of planetary systems, but because these environmental influences are likely the norm, and hence affect the majority of stars and planets in the Galaxy.\\

With the launch of JWST, it is now possible to observe distant, massive SFRs in unprecedented detail. The sources observed in this study reside within the giant Galactic massive SFR of NGC 3603, at a distance of $7.2 \pm 0.1$ kpc \citep{drew2019star}. With its compact central cluster of OB stars, NGC 3603 is the Milky Way's closest analogue to a starburst cluster. In order to efficiently study the PMS stars in this cluster, we have utilized the Near Infrared Spectrograph's (NIRSpec) multi-object spectroscopy (MOS) mode. NIRSpec's MOS mode makes use of the novel Micro-Shutter Assembly (MSA), an array of over $250\,000$ microscopic doors that can be configured opened or closed, allowing for the simultaneous observation of up to hundreds of astrophysical spectra \citep{ferruit2022near}. We have obtained spectra of 100 sources in high resolution mode using the $G235H-F170LP$ grating/filter combination.\\ 
In this paper, the spectra that we have classified as PMS stars are presented and discussed. In section ~\ref{sec:sel_of_tar} we discuss the target selection. In section ~\ref{sec:data_reduction}, the data reduction process is explained. In section \ref{post_pro}, we describe additional calibration and processing steps including the nebular subtraction procedure that we have developed. In section \ref{sec:methods}, we describe our methods including our spectral fitting procedure. In section ~\ref{sec:results}, we present our results including the spectral type, ages and masses and accretion rates of our sources. In section ~\ref{sec:discussion} these results are discussed. In section ~\ref{sec:conclusions} our findings are summarised.

\section{Selection of targets and observation strategy}
\label{sec:sel_of_tar}
The sources in this study were originally observed with Hubble Space Telescope's (HST) Wide Field Camera 3 (WFC3) (PID: 11360, PI: O'Connell, 2009). The filters employed in this observing programme were F555W, F625W, F814W, F110W and F160W. A different programme provided the F435W filter from HST/ACS observations (PID: 10602, PI: Maiz-Apellaniz, 2005). We have used these archival HST observations as well as our new NIRSpec observations to spectrally classify our sources (see section \ref{subsec:ext_corr}). In \citet{beccari2010progressive}, hereafter  B10, the F555W and F814W filters were used to characterise the PMS population of NGC 3603. These sources were classified as either MS or PMS based on photometric $H_{\alpha}$ excess emission (with equivalent width $EW \ge 10 \AA$), which is known to trace accretion. Approximately 10,000 sources were observed in total. Of these 10,000 sources, 100 were selected to be observed by NIRSpec, of which 60 had been classified as PMS based on $H_{\alpha}$ excess above the 5 $\sigma$ level. The remaining 40 did not show evidence of $H_{\alpha}$ excess above the 5 $\sigma$ level, and so were classified as MS stars. This strict selection criterion was employed in B10 to avoid the inclusion of sources with poorly subtracted nebular backgrounds or chromospheric emission masquerading as accretion, as both of these effects could lead to modest $H_{\alpha}$ excess emission. The approach of B10 was therefore highly conservative in identifying PMS stars. It is quite likely that many of the sources classified as MS are in fact PMS stars with intrinsically weak accretion signatures, or alternatively with significant NIR veiling, leading to a weaker detection of line emission. From our NIRSpec observations, we have classified $42$ sources as PMS stars, based on the presence of at least two of the strong hydrogen emission lines $Pa_{\alpha}$, $Br_{\gamma}$, and $Br_{\beta}$ in emission above the chromospheric level, after performing corrections for photospheric absorption and veiling. Of the 42 sources that we have spectroscopically classified as PMS, 25 were classified as PMS in B10. The remaining 17 were classified as MS. Of the 17 sources in which our classification differs to B10, some of these sources display weak accretion signatures which could have been overlooked due to the strict selection scheme in B10. Additionally, a number of them lie in regions of complex nebulosity, requiring a careful subtraction of the nebular emission in order to properly reveal the underlying accretion related emission (see section \ref{subsec:neb_sub}). However, there are a number of robustly accreting sources with bright emission lines in their NIRSpec spectra that had been classified as MS in B10. In these cases it is harder to reason why they may have been misclassified, though differences in methodology between this paper and B10 including background subtraction and EW measurement could play a role, along with intrinsic variability of PMS stars that strongly affects the strength of emission lines.\\
Compared to the sample used in our earlier work \citep{rogers2024determining}, the sources that we classify as PMS stars here include an additional 9 sources, bringing the total from $34$ to $42$. This is the result of improvements in the methodology allowing us to identify more sources with confidence, most importantly in the subtraction of the nebular background to reveal accretion related emission.\\

For our JWST NIRSpec observations, 3 micro-shutters were opened per source, forming a 'mini-slit'. The neighbouring shutters above and below the source shutter were opened in order to measure the nebular emission, and subtract it from the stellar spectra \citep[see also][]{demarchi2024}. This is a crucial procedure in a region like NGC 3603 where the nebular emission is bright. A 3 nod strategy was employed moving the source from the central shutter, to the upper shutter, and finally to the lower shutter. This pattern allowed for source spectra to be measured on different regions of the detector. The repeated exposures were eventually averaged together. This improved the S/N, while making it easier to remove cosmic rays, spurious pixels and other detector blemishes. The final observation strategy consisted of 3 unique MSA configurations, with 3 nods per configuration, resulting in 100 stellar spectra, and 600 nebular spectra.\\
In order to obtain a clean nebular spectrum for each target star, the micro-shutters observing the nebula must be free of stars. Due to the intense crowding of NGC 3603, this criterion placed tight constraints on which sources could be observed. The final list of targets was strongly influenced by how isolated each source was. Despite these precautions, it was not possible to completely avoid contamination of unwanted stars in nebular background micro-shutters. The number of usable nebular spectra decreased from $600$ to $471$ due to the presence of stars within the nebular micro-shutters. However, this was more than sufficient for the nebular subtraction.

\section{Data reduction}
\label{sec:data_reduction}
\subsection{NIPS}
\label{subsec:nips}
The uncalibrated data were largely reduced with the ESA Instrument Team’s pipeline known as the NIRSpec Instrument Pipeline Software (NIPS) \citep{NIPS}. Additional reduction steps were also written specifically for these data which are discussed below. NIPS is a framework for spectral extraction of NIRSpec data from the count-rate maps, performing all major reduction steps from dark current and bias subtraction to flat fielding, wavelength and flux calibration, background subtraction and extraction, with the final product being the 1D extracted spectrum.\\
One of the final steps before extraction is the rectification of the spectrum, which is necessary because the dispersed NIRSpec spectra are curved along the detector. Rectification is performed in order to 'straighten' the spectrum. By doing this, it also resamples the spectrum onto a uniform wavelength grid (each wavelength bin $\Delta \lambda$ is the same for every pixel). In our case, the rectified or ``regular 2D'' spectrum, was a count-rate map consisting of 3817 pixels in the dispersion direction and 7 pixels in the cross-dispersion direction.\\
The final data reduction step, namely the extraction, simply collapses the 2D spectrum by summing the 7 pixels along each column. Rather than working with the extracted 1D spectra, we chose to work with the rectified spectra, and performed further processing before eventual extraction in order to improve the S/N and aesthetic quality of the spectra.
\subsection{Optimal extraction}
\label{subsec:spec_avg_opt_extr}
During spectral extraction, a common approach is to simply sum over all of the pixels that the slit projects onto. In the case of a single micro-shutter, this is seven pixels along the detector column. For unresolved point sources, the majority of the flux is centred on just one or two detector rows. Using the simple sum extraction then leads to an unnecessary amount of noise being added to the spectrum, as most of the pixels contain little to no stellar light, but do contain detector noise. In order to minimise the noise that was added during the collapse of the spectrum, while including all pixels that contained stellar flux, an optimal extraction technique was employed, following the approach of \cite{1986PASP...98..609H}. This method used all seven pixels per column to extract the spectra, but inversely weighted pixels based on their noise level. Including all pixels ensures that all of the stellar light is considered, while suppressing the pixels that contributed mostly noise.\\ 
A quick check was done on the S/N improvement achieved with optimal extraction. Two sources were chosen, one bright source with a $V$ band magnitude of $16.2$, and a faint source with $V=26.7$. The two spectra were extracted using both the simple sum method, and the optimal extraction method. For the faint source, a region of the spectrum was defined that contained only continuum, and the S/N was measured to be 11.23. Using the optimal extraction technique, the signal-to-noise increased to 16.2, an improvement of $43\%$. For the bright source, the same approach yielded a signal-to-noise improvement of about $4\%$. This was not surprising, as optimal extraction was developed to suppress detector related noise. This is the dominant source of noise in faint spectra, and hence optimal extraction leads to a large improvement in those cases. Bright spectra are dominated by photon noise, and so optimal extraction has a less dramatic impact.\\ 
Finally, we combined the three nodded spectra for each source together. This was a multi-step process in order to again maximise the S/N while also removing spikes in the spectrum that had not been captured by the reduction pipeline. This procedure is described in detail in Appendix \ref{subsec:app_spec_avg}.
\section{Data corrections and calibrations} \label{post_pro}
\subsection{Absolute flux calibration}
\label{subsec:flux_cal}
NIRSpec spectra are expected to have an absolute flux calibration accuracy of $\sim 10 \%$ \citep{gordon2022james}. We investigated whether this was consistent with our observations by comparing our PMS spectra with the HST photometry of our sources. These observations are all available on the Hubble Legacy Archive (HLA). We used the F160W (H band) flux of each source to assess the flux calibration. We performed an interpolation from the shortest wavelength of our spectra at $1.66$ $\mu m$ to the pivot wavelength of the F160W filter at $1.5369$ $\mu m$. We did so by fitting either a 2D or 1D polynomial to the spectrum of each source. For the majority of our sources, we are observing the Rayleigh-Jeans tail of their spectrum, which is well approximated by a quadratic curve. For younger sources with flat or rising SEDs, the spectral shape around $1.66$ $\mu m$ is better approximated by a linear function, and so we used a 1D polynomial. We found that in order to bring the NIRSpec spectra into agreement with the F160W fluxes, a typical correction factor of just $1.02$ needed to be applied. The $1 \sigma$ variation of this scaling factor was $0.11$. This is consistent with the $10\%$ accuracy expected, indicating little to no variability in the continuum of our PMS source spectra between the HST observations taken in 2010 and the JWST observations in 2022.
\subsection{Nebular background subtraction}
\label{subsec:neb_sub}
\subsubsection{The need for a scaled nebular subtraction}
The subtraction of the nebular background from the source spectra proved to be a significant challenge. NIPS is capable of performing an automatic subtraction at detector level, in which it subtracts from the trace of the stellar spectrum, an equally sized trace containing nebular emission, as measured in the neighbouring micro-shutters. The performance of this subtraction routine was assessed in \citet{rogers2022quantifying} based on simulated NIRSpec data for nebular emission with spatially uniform brightness. In that study, we found that the subtraction resulted in a residual nebular flux of $0.8\%$ with a spread of $13\%$. After inspecting the real nebular spectra obtained with NIRSpec, we found that the nebular emission did in fact change in brightness by typically a few percent over the scale of a few micro-shutters (micro-shutter area = $0.46\arcsec \times 0.2\arcsec$). The typical difference in emission line flux for $Br_{7}$ between neighbouring nebular regions was found to be $7 \%$. In five out of 42 PMS sources, the corresponding nebular emission line fluxes differed by $\ge 20\%$ between adjacent micro-shutters. In order to account for the spatially variable nebulosity, a scaling procedure was developed to subtract the nebular light from the stellar spectrum. 
\subsubsection{Can He I act as a scaling metric?}
Following \cite{demarchi2024}, our brightness scaling procedure was based on the removal of the He I emission line doublet centred at $1.869 \mu m$. This method assumes that the nebula is the dominant source of helium emission. Excitation from extreme-UV photons from the central OB stars is likely the dominant mechanism to produce this emission line, given the high excitation potential of this transition $\sim$ 23 eV. This doublet lies in a wavelength range that is very challenging to observe from the ground due to telluric absorption. It has also not been covered by the majority of infrared space telescopes, apart from the NICMOS instrument on HST, and now JWST NIRSpec. As such, this helium doublet has not been well studied or discussed in the literature. We have examined the available archival T Tauri and protostar spectra observed with NIRSpec, in order to determine whether this line may have a significant contribution from the circumstellar environment. The JWST programmes that observed these sources are given in table \ref{archive_jwst_programs}, along with the name of the sources and the source type.
\begin{table}
\caption{Sources with archival JWST NIRSpec observations.}
\begin{tabular}{@{} l *3c @{}}
\toprule
 \multicolumn{1}{c}{programme}    & Source ID& Source type& \\ 
\midrule
 PID: 2104; PI: Harsono& IRAS-04365+2535&  Protostar\\
 PID: 1644; PI: Dougados& DG-TAU-B& Protostar\\
 PID: 1706; PI: Nisini& HH46&  Protostar\\
 PID: 1186; PI: Greene& SER-S68N&  Protostar\\
 PID: 1186; PI: Greene& SER-SMM-3&  Protostar\\
 PID: 1621; PI: Pascucci& HH30&  Edge-on disk\\
 PID: 1621; PI: Pascucci& TAU042021&  Edge-on disk\\
 PID: 1621; PI: Pascucci& IRAS04302&  Edge-on disk\\
 PID: 1621; PI: Pascucci& HKTAUB&  Edge-on disk\\
 PID: 1621; PI: Pascucci& FSTAUB&  Edge-on disk\\ \bottomrule
 \end{tabular}
 \label{archive_jwst_programs}
\end{table}
All of these sources are located in Taurus-Auriga and Ophiuchus, where nebular emission is negligible or entirely absent. As such, the presence of He I $1.869$ $\mu m$ would strongly suggest an origin in the circumstellar environment. We found that this He I line is absent in all protostar spectra, with an EW consistent with 0. In the case of the edge-on disks, we found weak He I emission in three of the five sources, with EWs $\le 1 \AA$. In the remaining two edge-on disks, the feature was again absent. Additionally, this feature does not appear in absorption except for massive stars of spectral type mid-B and earlier \citep{husser2013new}. A number of the youngest and brightest sources in our sample, with the strongest emission lines, show no He I $1.869$ $\mu m$ emission in their spectra, even before nebular subtraction. These sources are so bright that the nebular recombination lines cannot be detected above the continuum level of the star. The lack of He I $1.869$ $\mu m$ in these sources demonstrates that even when hydrogen lines are strongly in emission, one should not necessarily expect to detect correspondingly strong circumstellar He I $1.869$ $\mu m$. At the same time, it seems more likely to find He I $1.869$ $\mu m$ in more active sources, as this line requires either very high temperatures or high degrees of irradiation. As such, it is unlikely to be produced by weakly accreting sources.\\ 
Our nebular subtraction routine works by scaling the subtraction until this He I line has been fully removed from the stellar spectrum. To quantify the potential over-subtraction caused by removing the He I line, we have re-run the nebular subtraction for five sources with high accretion rates, where the line could plausibly form in the circumstellar environment. To begin, we ran the nebular subtraction as before, removing the He I completely. We then measured the resulting line flux for $Pa_{\alpha}$. We then performed the subtraction again, but stopped once the He I EW was reduced to $EW = 0.5 \AA$. The resulting change in $Pa_{\alpha}$ flux was modest, increasing on average by $9 \%$ when the He I EW was reduced to $0.5 \AA$.\\ 
These results together suggests that the circumstellar environment for PMS stars can, at best, produce only weak He I $1.869$ $\mu m$ emission, with EW at the level of $\le 1 \AA$. If this is the case for some of our more active sources, we may underestimate the accretion rate by of order $10\%$. More NIRSpec observations of CTTS and Herbig AeBe stars in low-mass star forming regions absent of nebular emission are needed to understand the extent to which He I $1.869$ $\mu m$ is produced in the circumstellar environment. For this study, we proceed by fully removing the line from our PMS spectra.\\

\subsubsection{The scaling procedure}
The procedure to scale our nebular subtraction worked as follows. The nebular spectrum was multiplied by a scaling factor, beginning with $0.01$. The scaled nebular spectrum was then subtracted from the source spectrum. The He I EW in the source spectrum was measured. This was repeated with the scaling factor increasing in steps of $0.01$ each time. The scaling factor that resulted in a He I EW as close to zero as possible was saved for that source, before moving on to the next source. In practice, after the scaled nebular subtraction, the stellar spectra had a typical He I EW of $0.011 \pm 0.16 \AA$. The typical scaling factor was $0.94 \pm 0.2$. The scaling factor being slightly below unity with low scatter further supports that the He I $1.869$ $\mu m$ is nebular. Sources producing significant circumstellar He I $1.869$ $\mu m$ would require a larger scaling factor, which would be $\gg 1$. Outliers such as this are not seen in our sample.\\
As a result of the nodding pattern employed for our observations, we obtained six nebular spectra for each stellar spectrum (though some of these were not useable due to contamination). This meant that we could normally attempt the nebular subtraction six times per source, using a different nebular spectrum each time. This was done because we found that for a given source, some nebular spectra resulted in a cleaner subtraction than others. Here we define ``clean'' as being a subtraction in which there was minimal residual in the He I line, and none of the hydrogen lines became negative due to over-subtraction. The variation in performance of different nebular spectra was likely due to the poor pixel sampling. NIRSpec does not Nyquist sample the line spread function (LSF) at any wavelength for point sources using the $R \; = \; 2700$ mode. In many cases weak emission lines were sampled by just 1 or 2 pixels. This meant that even a small difference in the distribution of flux across those pixels from the nebular spectrum to the stellar spectrum could be enough to cause over-subtraction in one of the pixels, resulting in a negative flux. Higher sampling in combination with higher spectral resolution would spread the flux over more pixels, making the subtraction less sensitive to this. Having access to multiple nebular spectra per source allowed us to mostly circumvent this problem, and typically we could find a nebular spectrum that resulted in a clean subtraction. We have employed the conservative approach that if a clean subtraction was not possible for a source, it was classified as MS.

\subsubsection{Does removing He I remove H I?}
\begin{figure*}[h]
    \centering
    \includegraphics[width=1\linewidth]{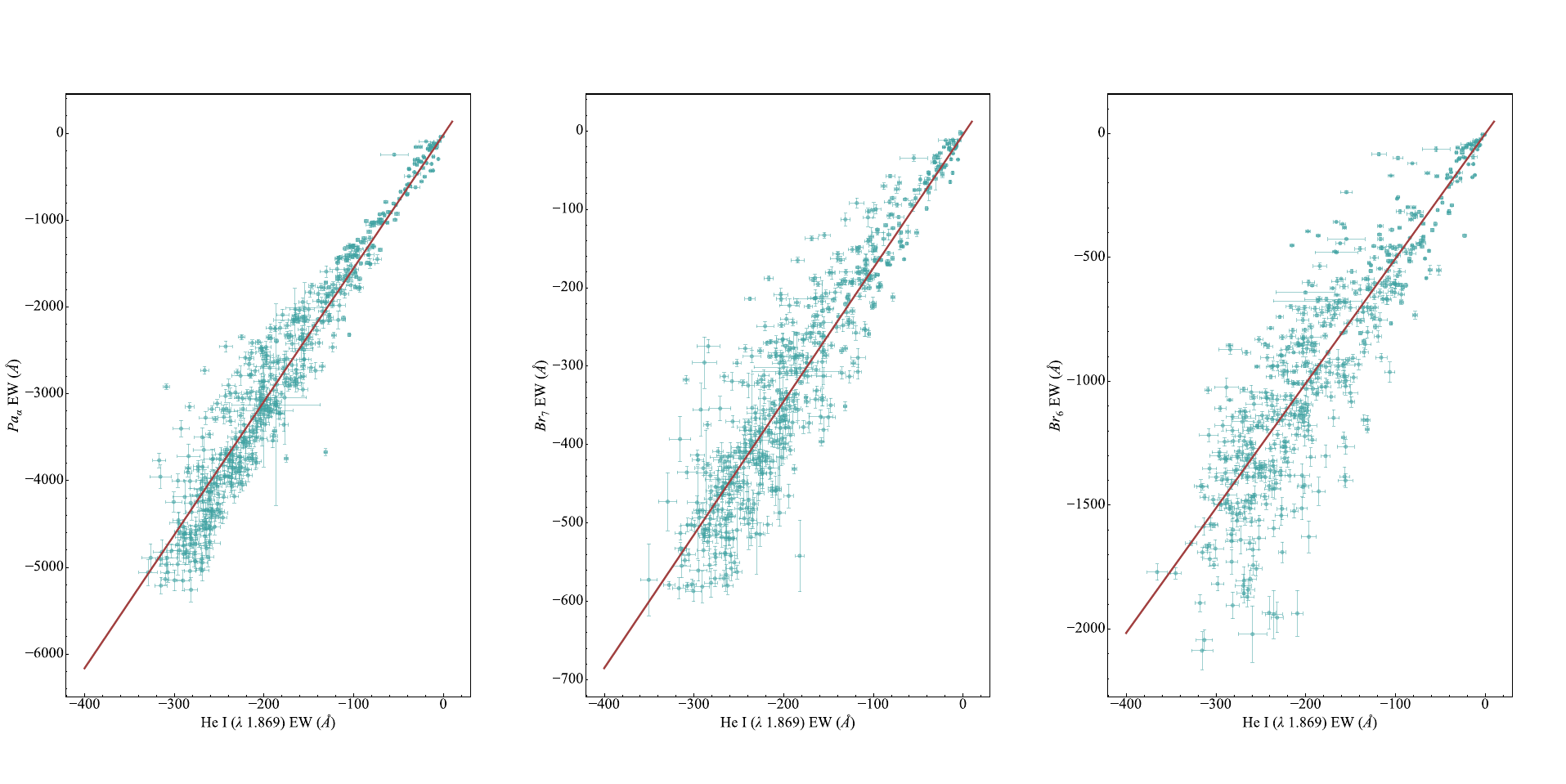}
    \caption{Relationship between the EW of He I $1.869 \mu m$ with: left, $Pa_{\alpha}$; centre, $Br_{7}$; and right, $Br_{6}$.}
    \label{fig:He_I_vs_recombination_lines}
\end{figure*}

Using the He I line as a metric for nebular subtraction relied on the assumption that the hydrogen recombination lines are also removed (i.e. there is a scaling relationship between nebular helium and hydrogen). In order to test the validity of this assumption, we measured the EW of the He I emission line, as well as the EW of the strongest H I emission lines: $Pa_{\alpha}$, $Br_{7}$ and $Br_{6}$ for all of our nebular spectra. The resulting relationship is shown in \ref{fig:He_I_vs_recombination_lines}, along with lines of best fit. All three hydrogen lines show a tight scaling relationship with He I. The dispersion around the line of best fit becomes larger for lines farther away in wavelength from He I. This implies that for longer wavelengths, other nebular lines may be needed to calibrate the nebular subtraction.\footnote{Indeed, to remove nebular emission from their lower-resolution ($R \; = \; 1000$) NIRSpec spectra of PMS stars in NGC\,346, \cite{demarchi2024} used both HeI lines at $1.869$ and $2.06$.}\\
Equations \ref{eq_1}, \ref{eq_2}, and \ref{eq_3} show the coefficients of the best fitting lines for $Pa_{\alpha}$, $Br_{7}$ and $Br_{6}$. The y-intercept of each equation indicates the EW of each hydrogen line when the EW of the He I line has been reduced to zero. In all three cases, some residual hydrogen emission is present according to these lines of best fit. A useful value to know is how much nebular flux may be left over in our subtracted stellar spectrum. In the case of $Pa_{\alpha}$, there are $19.95 \pm 2.26$ $\AA$ of hydrogen emission remaining. These EW values are of course with respect to the nebular continuum, which is significantly lower than the stellar continuum. In order to convert this EW to a flux, we have multiplied it by the typical continuum level of our nebular spectra. This equated to a typical residual flux of $4.29 \pm 0.49 \times 10^{-17}$ $erg/s/cm^2$. For context, the median $Pa_{\alpha}$ flux from our PMS sources was $2.03 \times 10^{-15}$ $erg/s/cm^2$. The residual nebular flux then corresponds to a typical under-subtraction at the level of $\sim 2\%$, which we do not deem significant with regard to the other uncertainties that affect our measurements. Given our argumentation in the previous section that we may even over-subtract some of our spectra, the true residual nebular flux is likely somewhere between $+2\%$ and $-8\%$ depending on the source in question.

\begin{equation} \label{eq_1}
    \centering
    Pa_{\alpha} = (15.36 \pm 0.026) \times He I - (19.95 \pm 2.26) \AA
\end{equation}

\begin{equation} \label{eq_2}
    \centering
    Br_{7} = (1.70 \pm 0.003) \times He I - (5.31 \pm 0.31) \AA 
\end{equation}

\begin{equation} \label{eq_3}
    \centering
    Br_{6} = (5.04 \pm 0.008) \times He I - (-3.03 \pm 0.77) \AA 
\end{equation}

\section{Methods}
\label{sec:methods}
\subsection{Determining spectral type, extinction and veiling}
\label{subsec:ext_corr}
In order to establish the age, mass and accretion rate of our sources, we needed to first determine their spectral type, as well as their extinction, and excess continuum flux related to the protoplanetary disk and accretion. This excess flux is referred to as ``veiling''. Veiling is usually defined as the ratio of disk emission to photospheric emission, denoted by $r_{\lambda}$. We implemented a Markov chain Monte Carlo (MCMC) exploration procedure to simultaneously determine the stellar and disk parameters by fitting our NIRSpec spectra and HST photometry with stellar models that had been both extinguished and veiled. Our method shares much in common with the approach outlined in \cite{herczeg2014optical} for optical spectra with moderate resolution. We also drew inspiration from \cite{starfish} with the incorporation of a MCMC exploration to obtain meaningful uncertainty estimates on each parameter. Our approach consisted of first fitting the normalised NIRSpec spectrum of each source in order to obtain an estimate of $T_{eff}$ and $r_{\lambda}$ based on the minimum $\chi^2$. These estimates would act as the basis of our priors for $T_{eff}$ and $r_{\lambda}$. For the remaining parameters we used uninformative (flat) priors. Following this, we employed MCMC in order to determine more precise parameter values and their uncertainties.\\
\subsubsection{Normalised spectrum fit}
We opted to use the Phoenix stellar models from \cite{husser2013new} as our starting point. We degraded the spectral resolution of the Phoenix models to match NIRSpec using a Gaussian convolution and resampling onto a common wavelength axis. These models have three parameters, namely, effective temperature $T_{eff}$, surface gravity $log(g)$ and metallicity $[Fe/H]$. We only considered a sub-sample of the Phoenix spectra, with $3000 < T_{eff} < 12000$, $log(g)$ = $4.00$ and $[Fe/H] = 0.0$. At the modest spectral resolution of NIRSpec, changes to absorption lines due to surface gravity are not significant, and so we elected to fix $log(g)$ to the typical value for CTTS of $4.00$ \citep[e.g.][]{herczeg2014optical}. The metallicity of NGC 3603 has been investigated in other studies and found to be consistent with $[Fe/H]=0.00$, and so we fixed it at this value. We review these studies briefly in section \ref{subsec:metallicity}. We normalised the NIRSpec spectra using a univariate spline function, setting the continuum level to unity. We ran a simple $\chi^2$ fitting routine on each normalised spectrum, with $3000 < T_{eff} < 12000$K and $0 < r_{\lambda} < 20$. The values of the best fitting model, and the standard deviation of the ten best fitting models were used to generate the normally distributed priors for $T_{eff}$ and $r_{\lambda}$ during the MCMC part of this analysis. 
\subsubsection{Extinction}
As our NIRSpec spectra cover a wavelength range that is not particularly sensitive to extinction, we also included HST photometry for each source. The filters employed were F435W (ACS), F555W, F625W, F814W, F110W, and F160W (WFC3). We performed aperture photometry for each source, accounting for the aperture correction and subtraction of the nebular background. This provided us with significantly better sensitivity to the extinction of each source. The optical fluxes are also sensitive to the photospheric continuum of the central stars, allowing for more accurate $T_{eff}$ determinations.\\ 
In a previous work in which we examined the extinction towards NGC 3603 based on nebular emission line decrements \citep{rogers2024spectral}, we used the NIR extinction curve parametrisation of \cite{fitzpatrick2009analysis} (FM09). This extinction curve worked well with our nebular spectra, providing a parameterisation with a variable exponent $\alpha$ that produced better fits to our observations than other extinction curves with a constant exponent. However, the FM09 extinction curve does not cover some of our bluest optical data points, where we are most sensitive to extinction. Opting for an extinction curve with unbroken coverage from the optical to the infrared, we found that the recent extinction curve of \cite{gordon2023one} (G23) resulted in good fits to our observations when we tested it in our MCMC routine. The value of $\alpha$ used for the NIR portion of this extinction curve was $\alpha=1.68467$, which is consistent with the typical value that we determined in \cite{rogers2024spectral}, within the uncertainties.

In the future, if the FM09 curve is extended to include bluer optical wavelengths, it would be of great interest to compare it to G23.\\
We extinguished the Phoenix model spectra using the G23 extinction curve, setting $R(V) = 4.8$, based on our results in \cite{rogers2024spectral}. We did not allow this parameter to vary so that the number of free parameters was kept to a minimum. $A(V)$ was permitted to vary between 0 and 10.
\subsubsection{Excess veiling emission}
Continuum emission from the protoplanetary disk can veil the stellar spectrum by reducing the apparent depth (i.e. the EW) of absorption lines. This emission is typically ascribed to hot dust at the dust sublimation radius of the protoplanetary disk \citep[e.g.][]{millan2006circumstellar}. One method to constrain NIR veiling in CTTS is to model the veiling emission as a blackbody with a single temperature ranging from $500 - 2000$\,K \citep[e.g.][]{muzerolle2003unveiling, cieza2005evidence}. This range represents the sublimation temperature for a variety of dust grain compositions and grain sizes expected in the protoplanetary disk. More recent works \citep[e.g.][]{antoniucci2017high, alcala2021giarps} argued that hot gas in the inner disk also contributes towards veiling. In those studies, the veiling emission could only be successfully modelled with a blackbody spectrum with temperatures $\ge 2000$\,K. In \cite{mcclure2013characterizing}, three blackbody temperatures were used to fit the NIR veiling from 1 to $4.5$ $\mu m$ of CTTS: a cool blackbody ranging from $500 - 2000$\,K to represent the dust; a warm blackbody from 2000\,K to the effective temperature of the star to represent optically thick gas in the inner disk; and finally the Rayleigh-Jeans tail of a hot blackbody of $\sim$ 8000\,K to represent the accretion shock luminosity. The spectral wavelength range of \cite{mcclure2013characterizing} is broader than our own, and more importantly goes to longer wavelengths where NIR veiling is stronger, justifying a multi-component fit to the veiling. For our observations we opted to fit a single blackbody, whose temperature could vary between 500\,K up to the best fitting effective temperature of the star based on the normalised fit. This approximates excess emission from both the dust sublimation radius as well as gas in the inner disk without providing too much flexibility to overfit our observations. The value of $r_{\lambda}$ was again allowed to vary between 0 and 20.
\subsubsection{Luminosity scaling term}
In order to bring the Phoenix spectra into agreement with our observations, a luminosity scaling term was introduced. This was simply a multiplicative term that we applied across the entire spectrum. This term was varied until good agreement was found between the luminosity of the Phoenix spectrum and the observed spectrum. In general this term varied between $10^{-23}$ and $10^{-21}$.
\subsubsection{MCMC exploration}
\begin{figure*}[htbp]
    \centering
    \begin{minipage}[b]{0.45\textwidth}
        \centering
        \includegraphics[width=\textwidth]{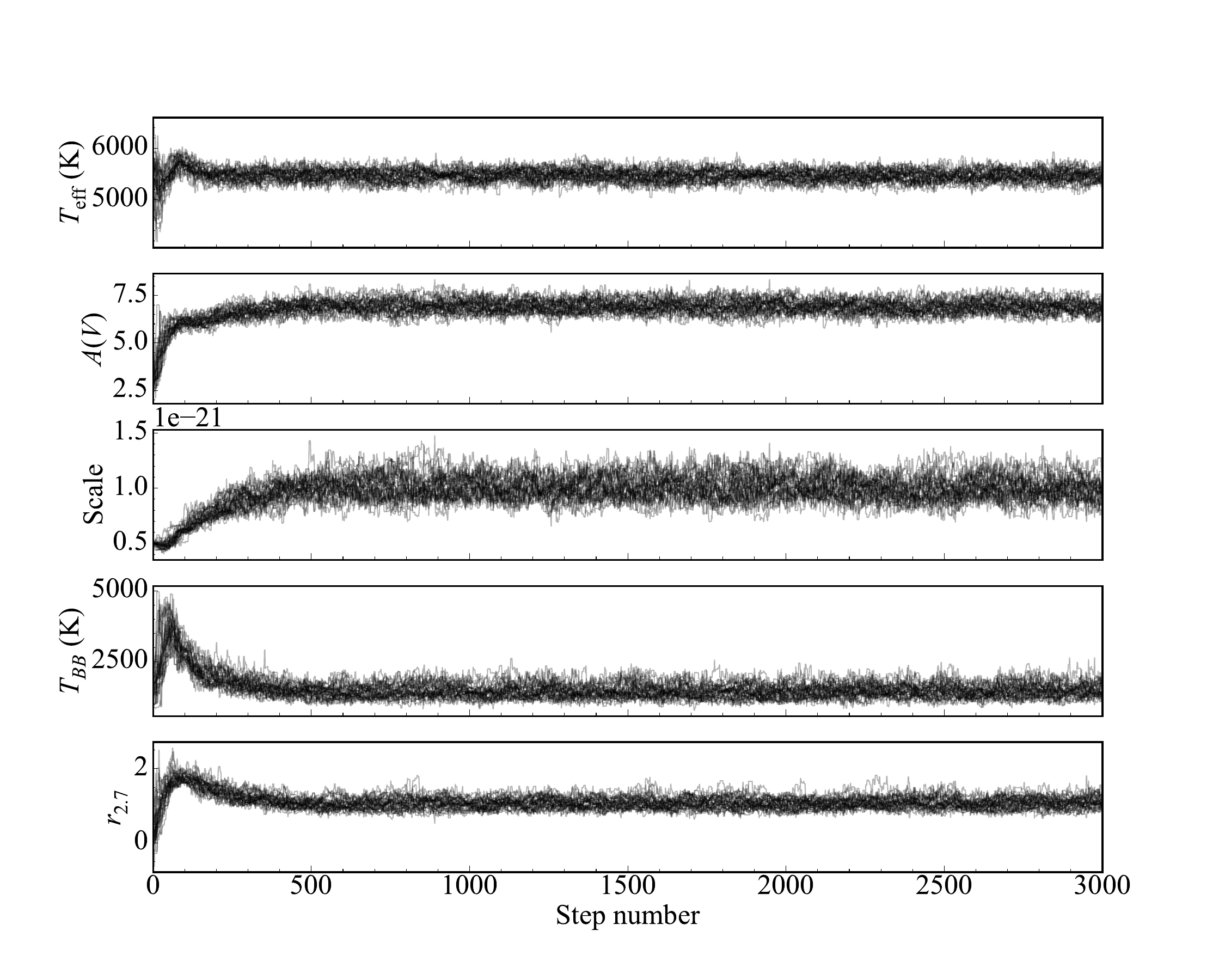}
        \caption{Trace plot of source 995.}
        \label{trace}
    \end{minipage}
    \hfill
    \begin{minipage}[b]{0.45\textwidth}
        \centering
        \includegraphics[width=\textwidth]{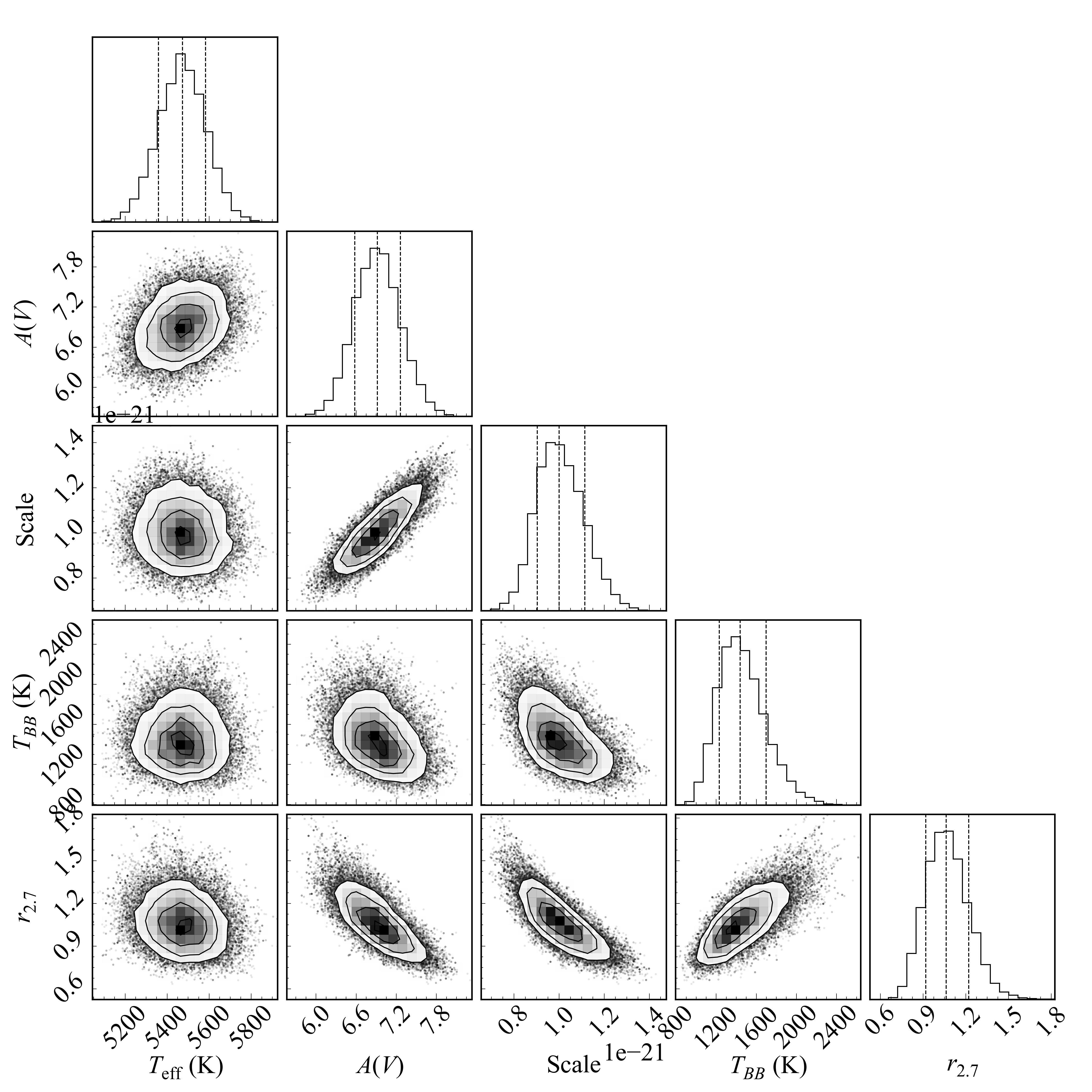}
        \caption{Corner plot showing the distribution of parameter values.}
        \label{contour}
    \end{minipage}
    
    \vspace{0.5cm} 
    
    \begin{minipage}[b]{1\textwidth}
        \centering
        \includegraphics[width=\textwidth]{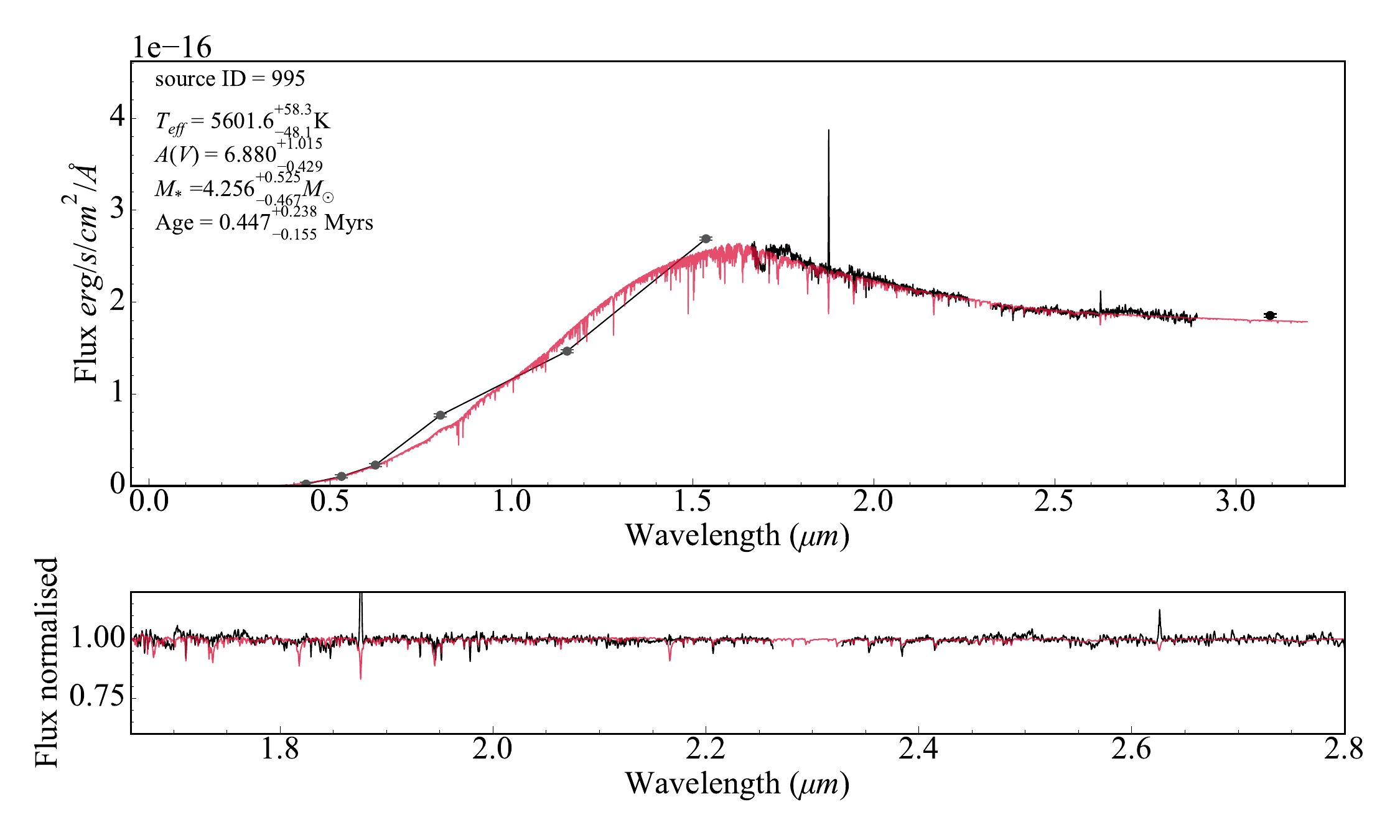}
        \caption{Best fitting model spectrum shown in red, overplotted on top of the source spectrum shown in black.}
        \label{best_fit_995}
    \end{minipage}
\end{figure*}

We employed a MCMC exploration with the python package emcee to determine realistic uncertainties on the best fitting parameter values. MCMC is an iterative sampling method to fit models to data. It works by generating a model based on the input parameters and randomly sampling values for each parameter from a pre-defined distribution (the priors). Each time a new model spectrum is generated, it is fitted to the observed spectrum, and the log-likelihood is calculated and taken as a goodness of fit metric.\\ 
If the current model provides a better fit than the previous model, then a new model is generated using the current parameter values as a starting point, plus a small random input. These small random steps in parameter space are accomplished by ``walkers''. If the current model is worse than before, then the next model is generated using the previous parameter values as the starting point, again, plus a small random input. This is repeated thousands of times until a reasonable range of parameter values has been determined, and no further exploration of the parameter space can improve the fit. This is known as convergence.\\ 
MCMC requires that arbitrarily small step sizes can be taken in parameter space when generating a new model to fit the observations. To enable this, we needed to interpolate the Phoenix model spectra. We used the radial basis function interpolation method from SciPy to do this. This method is computationally efficient, and works well for spectral features that change in a roughly linear fashion, which is a reasonable approximation for absorption lines with changing $T_{eff}$\\ 
We allowed the MCMC algorithm to run for a maximum of $10,000$ iterations per walker, utilising 30 walkers in total. This meant that for each iteration, 30 new models were generated and fitted to the observed spectrum. This parallelisation allows for a more efficient exploration of parameter space. We checked for convergence by comparing the autocorrelation time of each parameter with the number of iterations that had been completed. If the number of iterations was $\ge 50$ times the autocorrelation time, we stopped the procedure, as this indicates that the walkers had explored the parameter space sufficiently. Generally we reached convergence after approximately $3000 - 8000$ iterations.\\ 
Figure \ref{trace} shows an example of a trace plot produced by the fitting procedure. This plot visualises how the walkers have explored the parameter space and on which values they have converged. Figure \ref{contour} is a corner plot, illustrating the distribution of parameter values. Figure \ref{best_fit_995} shows the source spectrum and the best fitting model spectrum. The remaining sources and their best fitting model spectra are available \href{https://zenodo.org/records/15188940}{here}.

\subsection{Classifying continuum sources} \label{subsec:class_cont}
A number of sources in our sample displayed extremely weak or no absorption lines in their NIRSpec spectra. Most of these sources exhibit rich emission line spectra, with transitions from H I, CO, Fe II among others. Strong emission line spectra are typically an indication of high levels of activity related to accretion. This serves to both fill in absorption lines with emission, while also leading to stronger NIR veiling, reducing the EW of absorption lines. Most of the continuum sources in our sample exhibit Class I or flat type SEDs in the NIR (see section \ref{subsec:spectral_index}), which is also a sign of strong excess veiling emission from the protoplanetary disk, and suggests that the sources still retain substantial circumstellar material. Without photospheric absorption lines, we had to rely solely on fitting the SED of these sources in order to constrain the stellar parameters. The HST observations proved critical in this step. The stellar photosphere is well traced by optical photometry as the star's light dominates at wavelengths between $0.4$ and $0.8$ $\mu m$ \citep{bell2013pre}.\\

\subsubsection{Indirect clues about the spectral type of the continuum sources}
The majority of the continuum sources are highly luminous and display no absorption lines in their NIRSpec spectra. The high luminosity of these sources can be explained if they are of intermediate to high mass, extremely young, or both. The complete lack of absorption lines from metals can be explained by one of three possibilities. The first, and least likely, is that NGC 3603 has a significantly sub-solar metallicity. This would render metal absorption lines intrinsically weak. However, numerous studies have examined the metallicity of NGC 3603, and point towards a solar, or near solar composition \citep[e.g.][]{melnick1989galactic, lebouteiller2008chemical}. As such, we disregarded this possibility. Another option is that the metal absorption lines from the photosphere of our sources are so strongly veiled that they are no longer detectable. We have simulated how much veiling emission would be needed in order to render the strong metal absorption feature of Mg I at $\lambda=1.71$ $\mu m$ completely undetectable for a 4000\,K star, at the typical S/N of our continuum sources. We have assumed $T_{bb}\;=2000$K. This would require a veiling factor of $\ge 20$. At this level of veiling, the SED shape in the IR changes dramatically, as the blackbody spectrum of the veiling emission completely dominates over the stellar photosphere. We found that at this level of veiling, it was not possible to match the SED shape of our observations. As such, we do not favour extremely high veiling as an explanation. A Final explanation is that after $\sim\;4000$K, metal absorption lines become weaker with increasing effective temperatures. At solar metallicity, metal lines vanish from spectra with temperatures of $T_{\rm eff}\ge 7000$\,K \cite{husser2013new}. These hotter photospheres are dominated by hydrogen absorption lines which, for our sources, are entirely filled in by emission lines, leaving the spectra with no detectable absorption. Veiling certainly plays a role in weakening the absorption lines further, but it seems more reasonable that a combination of moderate to high veiling with high effective temperatures is the reason why no metal absorption is seen for these five sources.

\subsubsection{Fitting the SEDs of our sources}
We initially attempted to fit the SEDs of our sources with the Robitaille Young Stellar Object (YSO) SED models \citep{robitaille2017modular, richardson2024updated}. Aside from the simplest family of models which represent a naked stellar photosphere, the Robitaille models attempt to fit numerous parameters related to the star itself, as well as the disk, envelope, geometry and inclination. We found that we lacked the wavelength coverage at longer wavelengths to constrain these model parameters. Given our modest wavelength range from $0.4$ to 3 $\mu m$, we opted to only fit the optical HST photometry of our sources from $0.4$ to $0.8$ $\mu m$. By fitting only the optical observations, the underlying stellar properties could be determined without attempting to constrain the properties of the circumstellar environment. We performed a MCMC exploration, fitting the optical portion of our source spectra with the Phoenix stellar models, allowing $T_{eff}$, $A(V)$ and the luminosity scaling term to vary. The models that we fitted are purely photospheric, while the observations show clear signs of photospheric emission plus disk emission. This necessitated that the photospheric models should not be significantly in excess of the observations at wavelengths $\ge$ 1 $\mu m$, where NIR veiling emission becomes strong.\\ 
We achieved good fits for our continuum sources. The best fitting temperatures tend to be relatively hot, with a median value of 6547\,K compared to 4980\,K for the rest of the sample. 

\subsubsection{Comparing SED fitting results to optical spectra}
\begin{figure}[htbp]
    \centering
    \begin{minipage}[b]{0.45\textwidth}
        \centering
        \includegraphics[width=\textwidth]{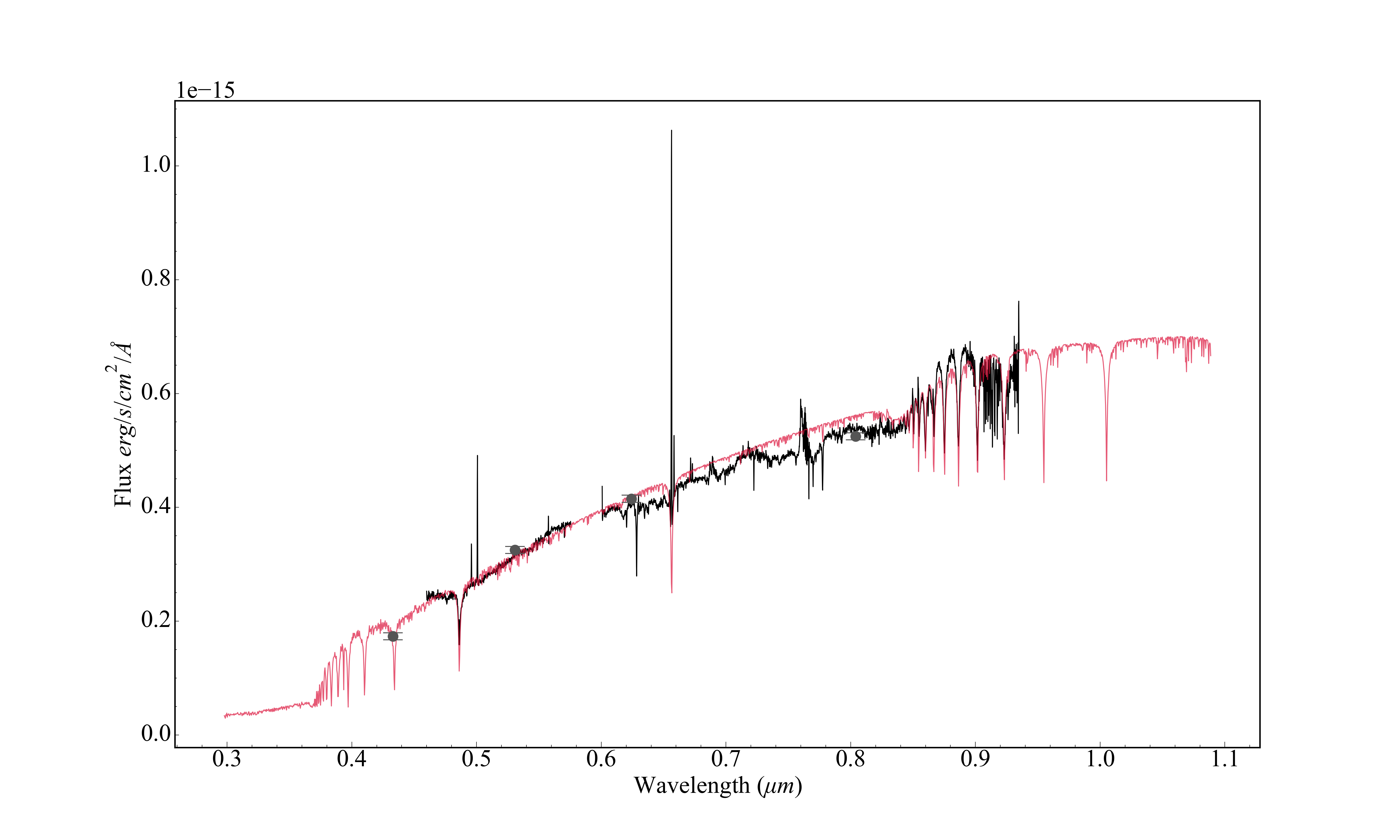}
        \caption{Best fitting optical model spectrum for source $152$ based on its MUSE spectrum. The black line corresponds to the MUSE source spectrum, the red line to the best fitting model spectrum, while the HST photometry is shown in grey.}
        \label{MUSE_152}
    \end{minipage}
    \vfill
    \begin{minipage}[b]{0.45\textwidth}
        \centering
        \includegraphics[width=\linewidth]{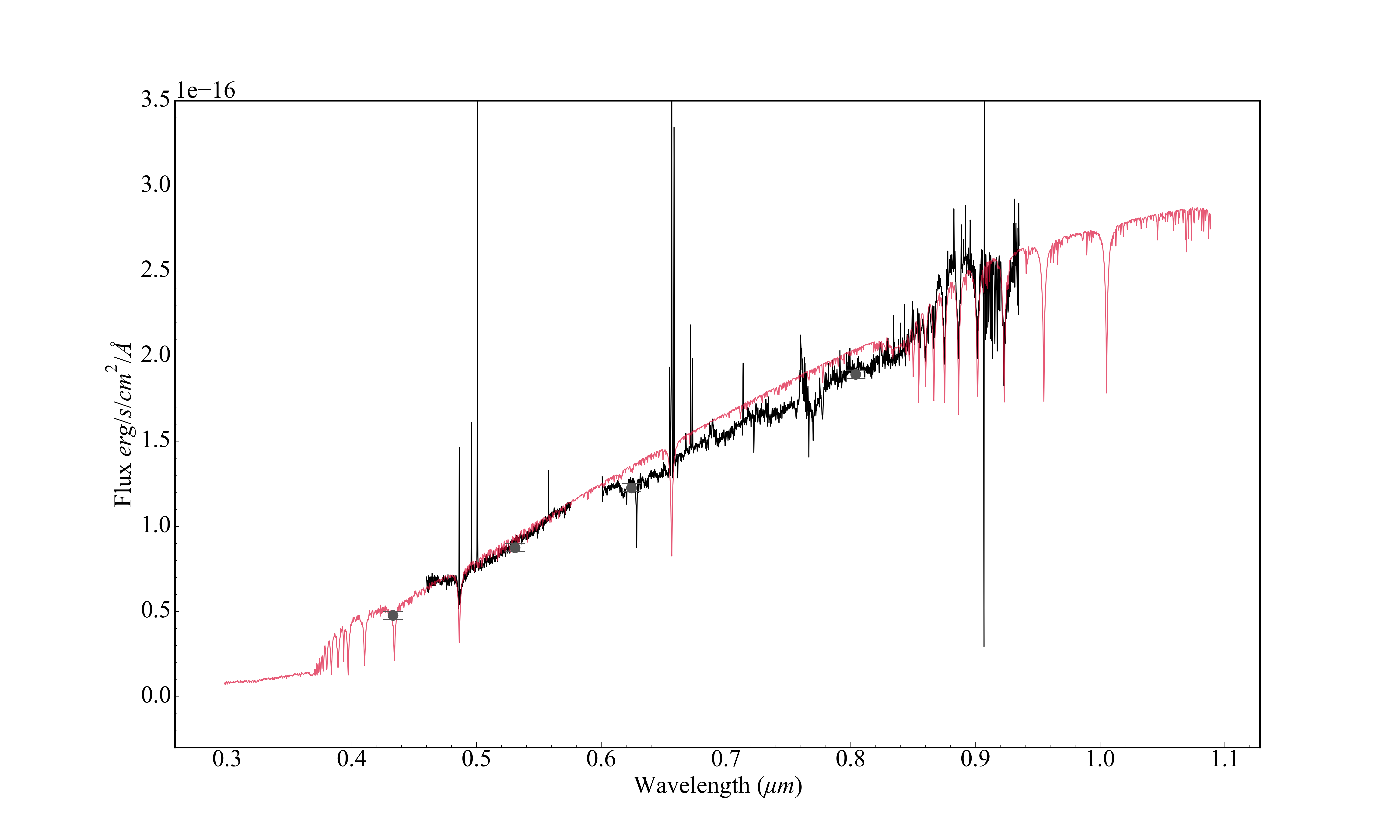}
        \caption{As above for source $354$.}
        \label{MUSE_354}
    \end{minipage}
\end{figure}

The central region of NGC 3603 was observed with MUSE by \cite{kuncarayakti2016unresolved}. A number of our sources fall within the field of view of those observations, providing us with a portion of their optical spectrum from $0.4$ to $0.9$ $\mu m$. Two of those sources, 354 and 152, are part of our continuum source sample, although they actually exhibit a single absorption line, $Br_{11-4}$. The presence and depth of this line indicates an effective temperature  $T_{\rm eff} \ge6000$\,K. The optical spectra of 354 and 152 are shown in figures \ref{MUSE_152} and \ref{MUSE_354}, respectively.
The MUSE spectra overlap well with each source's HST photometry. Despite strong emission lines in their optical spectra, a number of hydrogen absorption lines are still detected, including $H_{\beta}$, and a portion of the Paschen series. This provided us with the ability to compare the results of our SED fitting approach to a spectroscopic classification. We fitted the MUSE spectra of 354 and 152 with the Phoenix stellar models, allowing the temperature, extinction, and luminosity scaling term to vary, and calculating $\chi^2$ each time. The resulting best fitting model for 152 has an effective temperature of 7200\,K, compared to the SED best fitting temperature of 6185\,K. The best fitting spectroscopic temperature for 354 is 7000\,K, compared to 6875\,K from SED fitting.\\
This shows that we underestimated $T_{eff}$ in both cases, with the discrepancy being larger for source 152 than 354. What is clear from this exercise is that while our SED fitting results are more uncertain compared to a spectroscopic approach, we are more likely underestimating the true temperature of our sources, rather than overestimating it. This is consistent with the total lack of metal absorption lines in the continuum source spectra, which could only manifest for cool photospheres if an exceedingly high level of excess veiling emission were present. Our SED fitting shows that the continuum sources are relatively hot, intermediate mass sources, with a number of them belonging to the Herbig AeBe category. The stellar properties of these sources are shown in table \ref{tab:photospheric_properties_table}. 

\subsection{Measuring the recombination lines.}
\label{subsec:measure_lines}
To measure the recombination lines in the stellar spectra, we employed Monte Carlo simulations in order to propagate all of the uncertainties that arose from our processing steps and corrections. To do this, we generated $1000$ realisations of each recombination line, allowing the lines to vary within their uncertainties for each realisation. We fitted a Gaussian profile to each realisation of the recombination lines using the non-linear least squares fitting routine curve fit from SciPy. We calculated the EW of the best fitting Gaussian, and converted the EW to a flux by multiplying it by the adjacent flux calibrated continuum. The final flux of each line is the median of the $1000$ measurements. The uncertainty on the flux is the standard deviation of the $1000$ measurements. The nebular recombination lines were measured in the same way. In the appendix, figure \ref{fig:Pa_a_grid} shows the $Pa_{\alpha}$ emission line for all PMS sources, normalised to the continuum. The best fitting Gaussian profile is shown in each case as a red dashed line. 

\section{Results}
\label{sec:results}
In this section we discuss the physical properties of the sources, derived from their spectra. The final values of the effective temperature, luminosity, extinction, and veiling were determined using our MCMC procedure as described above. The results for each source are given in table \ref{tab:photospheric_properties_table}. 
\subsection{Spectral properties} \label{subsec_physical_properties}
We have converted effective temperatures to spectral types using the conversion scheme of \cite{pecaut2013intrinsic} for PMS stars. The majority of our sources (27/42) have spectral types G and F, with $5100 \le T_{eff} \le 6800$\,K. An additional 9 sources have type M and K with $3400 \le T_{eff} \le 5100$\,K. The remaining 6 sources have early F, A and late B types. The typical value of the extinction measured towards our sources is A(V) = $4.58$ with a physical spread of $\pm 1.85$. This is consistent with the typical foreground extinction of $A(V) = 4.5$ found in previous studies \citep[e.g.][]{melnick1989galactic, sung2004initial, melena2008massive}. The parameter that we were least able to constrain was the temperature of the NIR veiling blackbody emission. In many cases, the uncertainties are in the thousands of Kelvin. As we have discussed in section \ref{subsec:class_cont}, our lack of longer wavelength coverage means that our observations are not well suited to precisely determining NIR veiling parameters. We can in general conclude for a given source whether NIR veiling is present, but our ability to constrain the veiling properties beyond that is poor.

\begin{table*}
\caption{\small{Physical properties of PMS sources derived from our MCMC routine}}. 
\small
    \centering
    \begin{tabular}{ccccccccccc}
         \hline
         ID&  RA &  Dec &  $T_{eff}$ (K) & A(V) & SpT & $T_{bb}$ & $r_{2.7}$ & $M_*$ ($M_{\odot}$)& Age (Myr) \\
         \hline
         \hline
               523$^{a}$ & 168.77997 & -61.26359 & $4395.8^{+76.0}_{-83.9}$    & $3.736^{+0.1}_{-0.1}$ & K4 & - & -                                    & $1.299^ {+ 0.138 }_ {- 0.125 }$ & $0.104^ {+ 0.003 }_ {- 0.003 }$ \\
               185$^{b}$ & 168.76294 & -61.26247 & $7553.0^{+603.0}_{-607.0}$  & $5.821^{+0.3}_{-0.3}$ & A8 & - & $3.579\pm0.508$                      & $6.713^ {+ 0.777 }_ {- 0.696 }$ & $0.156^ {+ 0.067 }_ {- 0.047 }$ \\
               238$^{b}$ & 168.81542 & -61.25506 & $9893.0^{+656.0}_{-477.0}$  & $6.883^{+0.1}_{-0.1}$ & A0 & - & $0.706\pm0.130$                      & $6.876^ {+ 0.775 }_ {- 0.697 }$ & $0.171^ {+ 0.040 }_ {- 0.033 }$ \\
               251$^{b}$ & 168.76864 & -61.25309 & $5726.0^{+144.0}_{-125.0}$  & $5.235^{+0.2}_{-0.1}$ & G3 & - & $5.954\pm0.268$                      & $5.681^ {+ 0.191 }_ {- 0.185 }$ & $0.188^ {+ 0.031 }_ {- 0.027 }$ \\
               654 & 168.75432 & -61.25141 & $4267.1^{+56.9}_{-65.7}$    & $2.248^{+0.1}_{-0.1}$ & K4 & - & -                                    & $0.759^ {+ 0.060 }_ {- 0.056 }$ & $0.381^ {+ 0.085 }_ {- 0.069 }$ \\
               995 & 168.76336 & -61.24923 & $5601.6^{+58.3}_{-48.1}$    & $6.88^{+1.0}_{-0.4}$  & G4 & $1408.2 \pm 310$ & $1.397 \pm 0.440$     & $4.256^ {+ 0.525 }_ {- 0.467 }$ & $0.447^ {+ 0.238 }_ {- 0.155 }$ \\
               823$^{b}$ & 168.76255 & -61.27098 & $8967.0^{+352.0}_{-379.0}$  & $8.087^{+0.1}_{-0.1}$ & A2 & - & $5.626\pm0.304$                      & $5.020^ {+ 0.215 }_ {- 0.207 }$ & $0.487^ {+ 0.080 }_ {- 0.069 }$ \\
               469$^{b}$ & 168.79435 & -61.27794 & $6220.0^{+105.0}_{-105.0}$  & $5.555^{+0.1}_{-0.1}$ & F6 & - & $13.111\pm1.135$                     & $4.367^ {+ 0.132 }_ {- 0.128 }$ & $0.548^ {+ 0.061 }_ {- 0.055 }$ \\
               152$^{b}$ & 168.78192 & -61.26485 & $6182.0^{+258.0}_{-164.0}$  & $3.533^{+0.2}_{-0.2}$ & F7 & - & $7.188\pm0.849$                      & $4.365^ {+ 0.265 }_ {- 0.250 }$ & $0.548^ {+ 0.151 }_ {- 0.118 }$ \\
               354$^{b}$ & 168.77063 & -61.26260 & $6875.0^{+137.0}_{-132.0}$  & $4.842^{+0.1}_{-0.1}$ & F1.5 & - & $2.058\pm0.068$                    & $3.811^ {+ 0.148 }_ {- 0.143 }$ & $0.910^ {+ 0.111 }_ {- 0.099 }$ \\
              1442 & 168.81232 & -61.26714 & $5129.2^{+268.0}_{-236.9}$  & $5.465^{+0.4}_{-0.4}$ & G9 & $3688.346 \pm 1527$ & $0.263 \pm 0.078$  & $2.114^ {+ 0.513 }_ {- 0.413 }$ & $0.913^ {+ 1.176 }_ {- 0.514 }$ \\
              1354$^{b}$ & 168.75209 & -61.26452 & $5509.0^{+521.0}_{-390.0}$  & $6.108^{+0.5}_{-0.4}$ & G5 & - & $9.792\pm1.371$                      & $2.527^ {+ 0.755 }_ {- 0.581 }$ & $0.949^ {+ 1.808 }_ {- 0.622 }$ \\
              5612 & 168.76002 & -61.24627 & $3635.6^{+64.8}_{-61.1}$    & $4.569^{+0.4}_{-0.4}$ & M1 & - & -                                    & $0.420^ {+ 0.045 }_ {- 0.041 }$ & $1.249^ {+ 1.340 }_ {- 0.647 }$ \\
               977 & 168.80948 & -61.26228 & $5807.1^{+58.9}_{-56.8}$    & $5.805^{+0.2}_{-0.2}$ & G2.5 & $3098.731 \pm 1147$ & $0.822 \pm 0.187$& $3.107^ {+ 0.216 }_ {- 0.202 }$ & $1.257^ {+ 0.278 }_ {- 0.228 }$ \\
              1066 & 168.77240 & -61.24978 & $3655.2^{+38.9}_{-40.2}$    & $1.226^{+0.1}_{-0.1}$ & M1 & - & -                                    & $0.438^ {+ 0.025 }_ {- 0.024 }$ & $1.268^ {+ 0.270 }_ {- 0.223 }$ \\
              1350$^{b}$ & 168.80333 & -61.24829 & $11611.0^{+273.0}_{-634.0}$ & $8.615^{+0.0}_{-0.0}$ & B8.5 & - & $1.319\pm0.123$                    & $3.379^ {+ 0.044 }_ {- 0.043 }$ & $1.715^ {+ 0.079 }_ {- 0.075 }$ \\
               996 & 168.79757 & -61.26405 & $5623.6^{+55.0}_{-53.7}$    & $4.662^{+0.2}_{-0.2}$ & G4 & $4750.416 \pm 1107$ & $1.96 \pm 0.269$   & $2.365^ {+ 0.177 }_ {- 0.165 }$ & $2.302^ {+ 0.532 }_ {- 0.432 }$ \\
              2673 & 168.79780 & -61.26489 & $5498.7^{+58.4}_{-59.3}$    & $5.705^{+0.5}_{-0.7}$ & G5 & $4857.044 \pm 1251$ & $2.385 \pm 1.172$  & $2.215^ {+ 0.500 }_ {- 0.408 }$ & $2.669^ {+ 2.199 }_ {- 1.206 }$ \\
              1038$^{b}$ & 168.79597 & -61.26429 & $5699.8^{+466.0}_{-466.0}$  & $4.799^{+0.5}_{-0.5}$ & G3.5 & -  & $1.208\pm4.276$                   & $1.996^ {+ 0.524 }_ {- 0.415 }$ & $2.725^ {+ 4.710 }_ {- 1.726 }$ \\
               727 & 168.80987 & -61.26355 & $3709.8^{+20.3}_{-21.7}$    & $0.002^{+0.0}_{-0.0}$ & M0.5 & - & -                                  & $0.504^ {+ 0.017 }_ {- 0.017 }$ & $2.782^ {+ 0.242 }_ {- 0.223 }$ \\
               853 & 168.76448 & -61.24880 & $6127.4^{+57.7}_{-53.1}$    & $4.871^{+0.2}_{-0.3}$ & F7 & $4949.411 \pm 860$ & $4.485 \pm 0.847$   & $2.392^ {+ 0.168 }_ {- 0.157 }$ & $2.947^ {+ 0.634 }_ {- 0.522 }$ \\
              2122 & 168.80313 & -61.26121 & $5209.6^{+54.6}_{-58.9}$    & $5.501^{+0.3}_{-0.2}$ & G8 & - & -                                    & $1.854^ {+ 0.150 }_ {- 0.139 }$ & $2.955^ {+ 1.229 }_ {- 0.868 }$ \\
              1997 & 168.80546 & -61.25098 & $5579.4^{+57.4}_{-54.6}$    & $6.103^{+0.4}_{-0.4}$ & G4.5 & $4828.715 \pm 1337$ & $0.972 \pm 0.387$& $2.080^ {+ 0.450 }_ {- 0.370 }$ & $3.093^ {+ 1.922 }_ {- 1.185 }$ \\
              2166 & 168.79462 & -61.26612 & $5850.7^{+59.9}_{-57.1}$    & $6.359^{+0.6}_{-0.6}$ & G2 & $4799.778 \pm 1108$ & $4.070 \pm 0.61$   & $1.934^ {+ 0.460 }_ {- 0.372 }$ & $4.545^ {+ 3.529 }_ {- 1.987 }$ \\
               862 & 168.74841 & -61.25478 & $5910.3^{+57.3}_{-54.1}$    & $3.884^{+0.1}_{-0.1}$ & G1.5 & - & -                                  & $1.872^ {+ 0.072 }_ {- 0.069 }$ & $5.118^ {+ 0.620 }_ {- 0.553 }$ \\
               892 & 168.73242 & -61.25072 & $5488.4^{+56.8}_{-55.4}$    & $3.221^{+0.1}_{-0.1}$ & G5 & - & -                                    & $1.753^ {+ 0.079 }_ {- 0.076 }$ & $5.173^ {+ 1.046 }_ {- 0.870 }$ \\
              1271 & 168.73311 & -61.26350 & $5479.2^{+60.0}_{-60.0}$    & $3.644^{+0.1}_{-3.7}$ & G5 & - & -                                    & $1.503^ {+ 0.090 }_ {- 0.085 }$ & $7.772^ {+ 1.303 }_ {- 1.116 }$ \\
               871 & 168.82315 & -61.25514 & $6381.4^{+57.7}_{-55.8}$    & $4.047^{+0.1}_{-0.1}$ & F5.5 & - & -                                  & $1.643^ {+ 0.041 }_ {- 0.040 }$ & $8.019^ {+ 0.524 }_ {- 0.492 }$ \\
              1103 & 168.81884 & -61.26281 & $5482.3^{+62.9}_{-57.3}$    & $3.302^{+0.1}_{-0.1}$ & G5 & - & -                                    & $1.449^ {+ 0.068 }_ {- 0.065 }$ & $8.429^ {+ 1.175 }_ {- 1.031 }$ \\
              1436 & 168.73955 & -61.26267 & $3682.9^{+50.4}_{-49.8}$    & $0.857^{+0.2}_{-0.2}$ & M1 & - & -                                    & $0.510^ {+ 0.049 }_ {- 0.045 }$ & $8.845^ {+ 3.891 }_ {- 2.702 }$ \\
              1530 & 168.83358 & -61.26385 & $5660.5^{+55.4}_{-54.3}$    & $4.309^{+0.2}_{-0.2}$ & G4 & - & -                                    & $1.453^ {+ 0.101 }_ {- 0.094 }$ & $9.110^ {+ 1.950 }_ {- 1.606 }$ \\
               858 & 168.74497 & -61.26299 & $7423.4^{+98.1}_{-83.8}$    & $4.144^{+0.1}_{-0.1}$ & A9V & - & -                                   & $1.682^ {+ 0.056 }_ {- 0.054 }$ & $9.351^ {+ 0.525 }_ {- 0.497 }$ \\
              2880 & 168.82895 & -61.25743 & $4432.5^{+280.3}_{-272.1}$  & $3.299^{+0.6}_{-0.6}$ & K3.5 & - & -                                  & $0.857^ {+ 0.175 }_ {- 0.145 }$ & $9.367^ {+ 10.906 }_ {- 5.039 }$ \\
              1339 & 168.82001 & -61.26301 & $5273.4^{+57.1}_{-56.2}$    & $3.147^{+0.1}_{-0.2}$ & G7 & - & -                                    & $1.245^ {+ 0.078 }_ {- 0.074 }$ & $11.316^ {+ 2.486 }_ {- 2.038 }$ \\
              1717 & 168.81406 & -61.25073 & $5805.0^{+55.3}_{-56.9}$    & $4.619^{+0.2}_{-0.2}$ & G2.5 & - & -                                  & $1.348^ {+ 0.103 }_ {- 0.095 }$ & $11.694^ {+ 2.630 }_ {- 2.147 }$ \\
              1876 & 168.74696 & -61.26490 & $5143.8^{+54.9}_{-59.3}$    & $3.621^{+0.2}_{-0.2}$ & G9 & - & -                                    & $1.157^ {+ 0.111 }_ {- 0.102 }$ & $12.277^ {+ 4.019 }_ {- 3.028 }$ \\
              4104 & 168.80185 & -61.25822 & $5195.8^{+101.9}_{-59.7}$   & $5.242^{+0.9}_{-0.8}$ & G8 & $3377.5 \pm 1420$ & $3.870 \pm 0.499$    & $1.101^ {+ 0.246 }_ {- 0.201 }$ & $12.315^ {+ 8.150 }_ {- 4.904 }$ \\
              1981 & 168.81381 & -61.24985 & $5080.1^{+412.3}_{-331.9}$  & $3.54^{+0.6}_{-0.4}$  & G9.5 & - & -                                  & $1.008^ {+ 0.160 }_ {- 0.138 }$ & $12.993^ {+ 10.616 }_ {- 5.842 }$ \\
              1852 & 168.81640 & -61.25061 & $3316.9^{+55.1}_{-51.6}$    & $0.11^{+0.2}_{-0.3}$  & M3 & - & -                                    & $0.261^ {+ 0.037 }_ {- 0.033 }$ & $13.482^ {+ 4.535 }_ {- 3.393 }$ \\
              1854 & 168.72934 & -61.26002 & $5031.9^{+54.1}_{-57.4}$    & $3.026^{+0.3}_{-0.3}$ & K0 & - & -                                    & $1.065^ {+ 0.135 }_ {- 0.120 }$ & $14.112^ {+ 5.050 }_ {- 3.719 }$ \\
              1497 & 168.76901 & -61.25124 & $6390.0^{+56.4}_{-56.7}$    & $4.594^{+0.1}_{-0.1}$ & F5.5 & - & -                                  & $1.320^ {+ 0.032 }_ {- 0.031 }$ & $15.160^ {+ 1.258 }_ {- 1.162 }$ \\
              1813 & 168.80395 & -61.24603 & $6299.4^{+59.5}_{-55.3}$    & $4.516^{+0.2}_{-0.2}$ & F6 & - & -                                    & $1.206^ {+ 0.031 }_ {- 0.030 }$ & $19.436^ {+ 0.929 }_ {- 0.886 }$ \\
    \end{tabular}
    \tablefoot{\small{The sources are ordered in terms of increasing age.\\
    Source marked with $^a$ exhibits central inversion in emission lines and the determined $L_{acc}$ is a lower limit.\\ 
    Sources marked with $^b$ show no photospheric absorption features and their properties were determined from SED fitting alone.}}
    \label{tab:photospheric_properties_table}
\end{table*}

\subsection{Spectral index}
\label{subsec:spectral_index}
\begin{figure}[!hbt]
    \centering
    \includegraphics[width=1\linewidth]{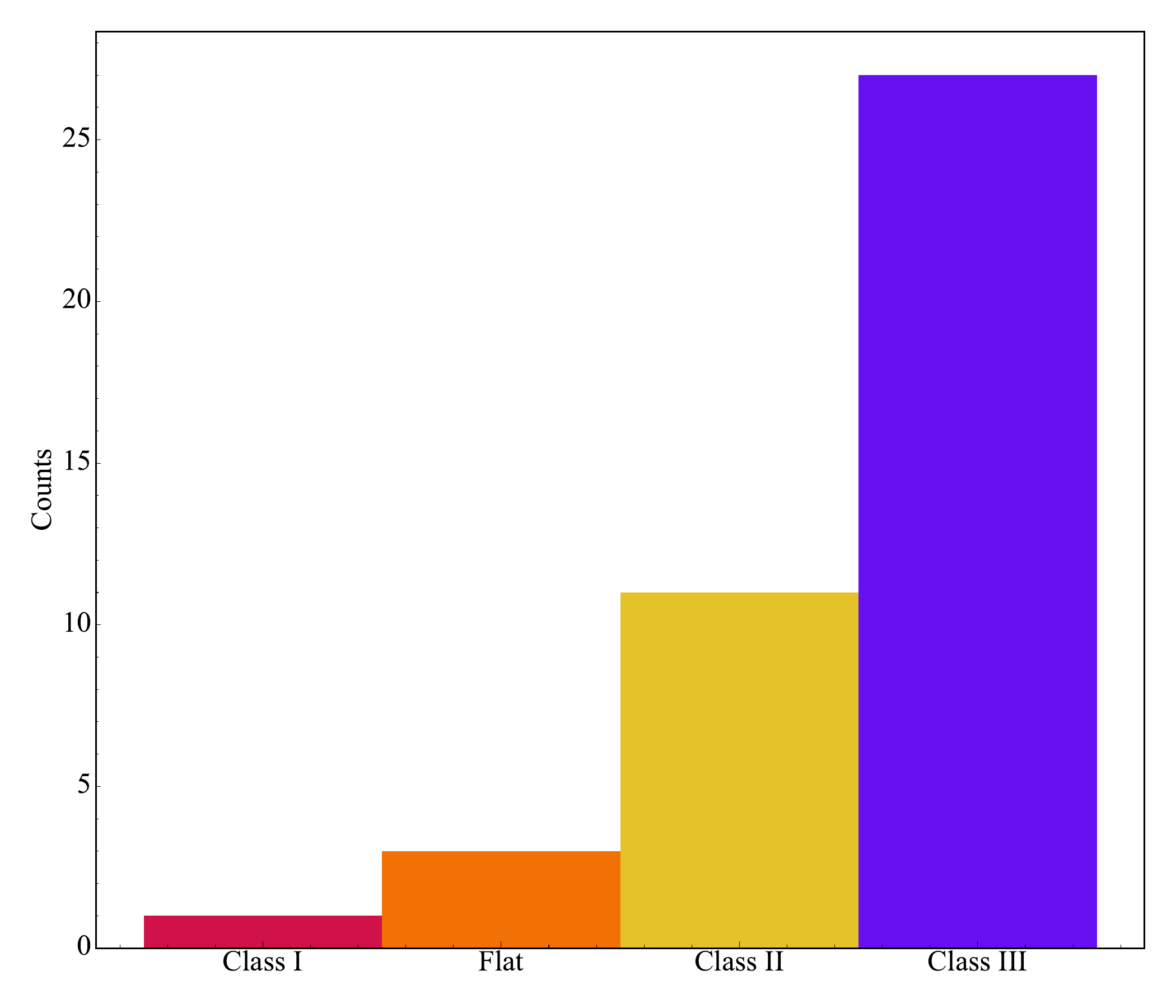}
    \caption{Histogram of evolutionary classes for our sample.}
    \label{class_hist}
\end{figure}

After correcting the sources for extinction, we measured the spectral index $\alpha$ of their NIRSpec spectra, defined as,
\begin{equation} \label{eq_alpha}
    \alpha = \frac{d \; log \; \lambda F_{\lambda}}{d \; log \; \lambda}
\end{equation} 
from \cite{wilking1989preprint}. This index is traditionally used to classify the evolutionary stage of PMS stars (i.e. Class I, II, III), with stars at earlier stages of evolution having a lower class. In \cite{greene1996near}, the authors defined as Class I sources those with $\alpha > 0.3$, as flat spectrum sources those with $0.3 \ge \alpha \ge -0.3$, as Class II sources as those with $-0.3 \ge \alpha \ge -1.6$, and as Class III sources those with $\alpha < -1.6$. We classified our sources using the same scheme. The resulting histogram is shown in figure \ref{class_hist}. The majority of our sources (26/42) belong to Class III. 
We point out that, typically, the index $\alpha$ is measured in the range $1-10\,\mu m$ using spectroscopy and/or photometry. Our wavelength range is therefore somewhat narrow, and we may underestimate the slope of our sources. CTTS SEDs in the IR are characterised by disk emission and dominated by hot dust longwards of 3 $\mu m$, which results in a rising SED. Without access to these longer wavelengths where dust really ``kicks in'' \citep[e.g.][]{furlan2011spitzer}, we may be measuring a local minimum in the SED of our Class III sources. With this caveat in mind, it is interesting to find that the majority of our sources lack detectable continuum emission from their protoplanetary disk at $2-3$ $\mu m$ (the $H$ and $K$ bands). Class III sources are thought to be essentially devoid of circumstellar material \citep{andre1994t}, so our Class III sources would go undetected as disk-bearing PMS stars using only $JHK$ broadband filter methods. However, the strong hydrogen emission lines in their spectra signal active and ongoing accretion, which necessitates that a gaseous disk still exists around the star. Accreting stars without detectable $JHK$ excess have been observed in other Galactic massive star forming regions. In their study on the pre-main-sequence population of Trumpler 37 within the Cepheus OB2 association, \cite{sicilia2005cepheus} reported that only $50\%$ of their accreting stars showed excess emission in their $JHK$ 2MASS photometry, despite showing strong $H_{\alpha}$ emission in their optical spectra. The authors go on to suggest external photoevaporation as a possible reason for this. We discuss this as a possibility in section \ref{subsec:ext_photoevap}.
\subsection{Stellar ages, masses, and radii}
\label{subsec:age_mass_radius}
\begin{figure*}[!hbt]
    \centering
    \includegraphics[width=1\linewidth]{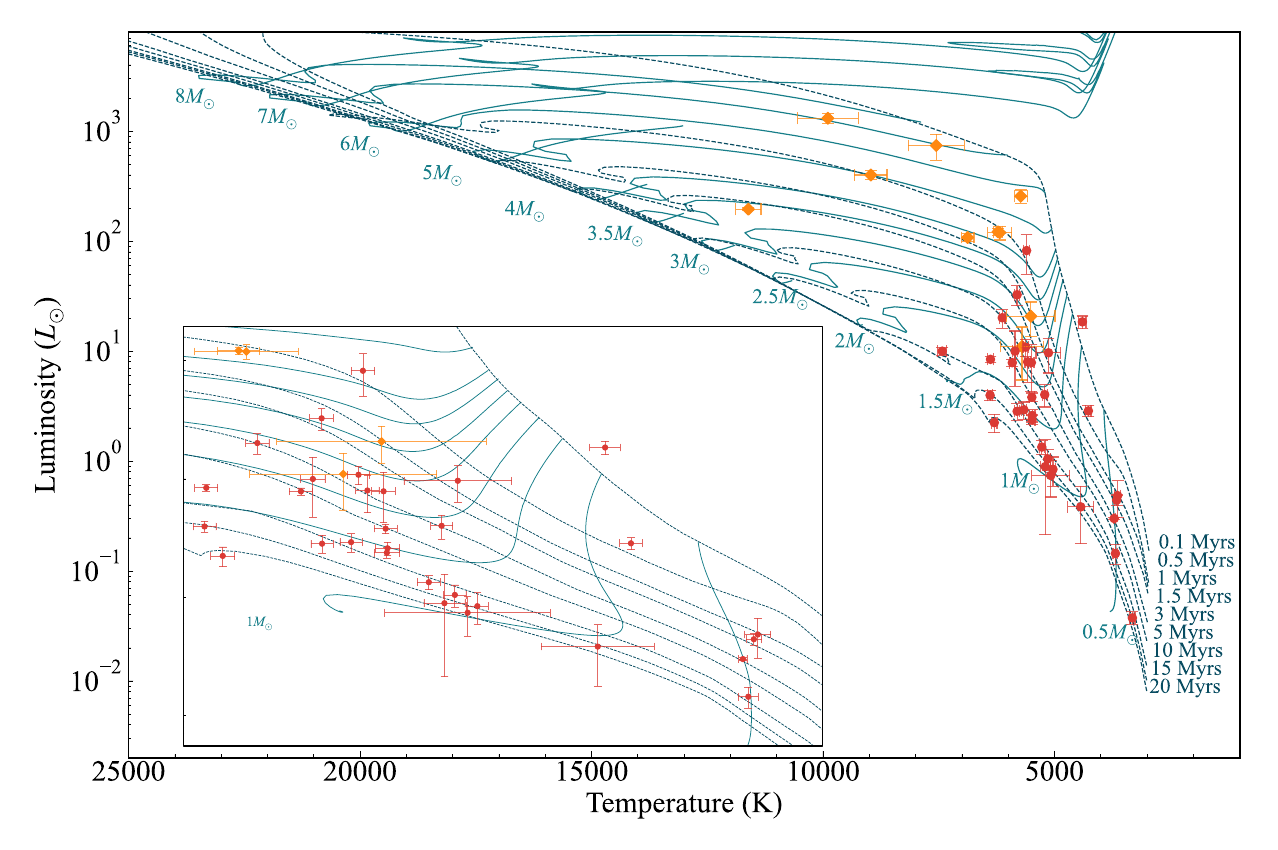}
    \caption{HRD of the PMS sources using spectroscopically determined $T_{eff}$ and $L_{*}$. The red points were classified spectroscopically. The orange diamonds were classified based on SED fitting. The 1$\sigma$ uncertainties are represented with red error bars. A selection of isochrones are shown as turquoise dashed lines. The ages are given in Myr on the right hand side.}
    \label{fig:hrd}
\end{figure*}

In order to derive more information about the physical properties of our sources, we placed them in the Hertzsprung--Russel diagram (HRD), shown in figure \ref{fig:hrd}. We used the MIST stellar isochrones and evolutionary tracks of \cite{choi2016mesa} in order to determine their ages, masses and radii. The uncertainties on these parameters were determined via Monte Carlo error propagation by taking the $16^{th}$ and $84^{th}$ percentiles using a log-normal distribution to account for the fact that the age mass and radius should always be positive. The majority of the sources have ages between $0.1$ to $3$ Myr, typical for accreting CTTS. However, we also report a sub-sample of twelve stars consistent with ages $\ge 10$ Myr. Of those, four are consistent with ages $\ge 15$ Myr. Stellar masses and ages are listed in \ref{tab:photospheric_properties_table}.\\ 
\subsection{Location of accreting sources within the cluster}
\begin{figure}[h]
    \centering
    \includegraphics[width=0.85\linewidth]{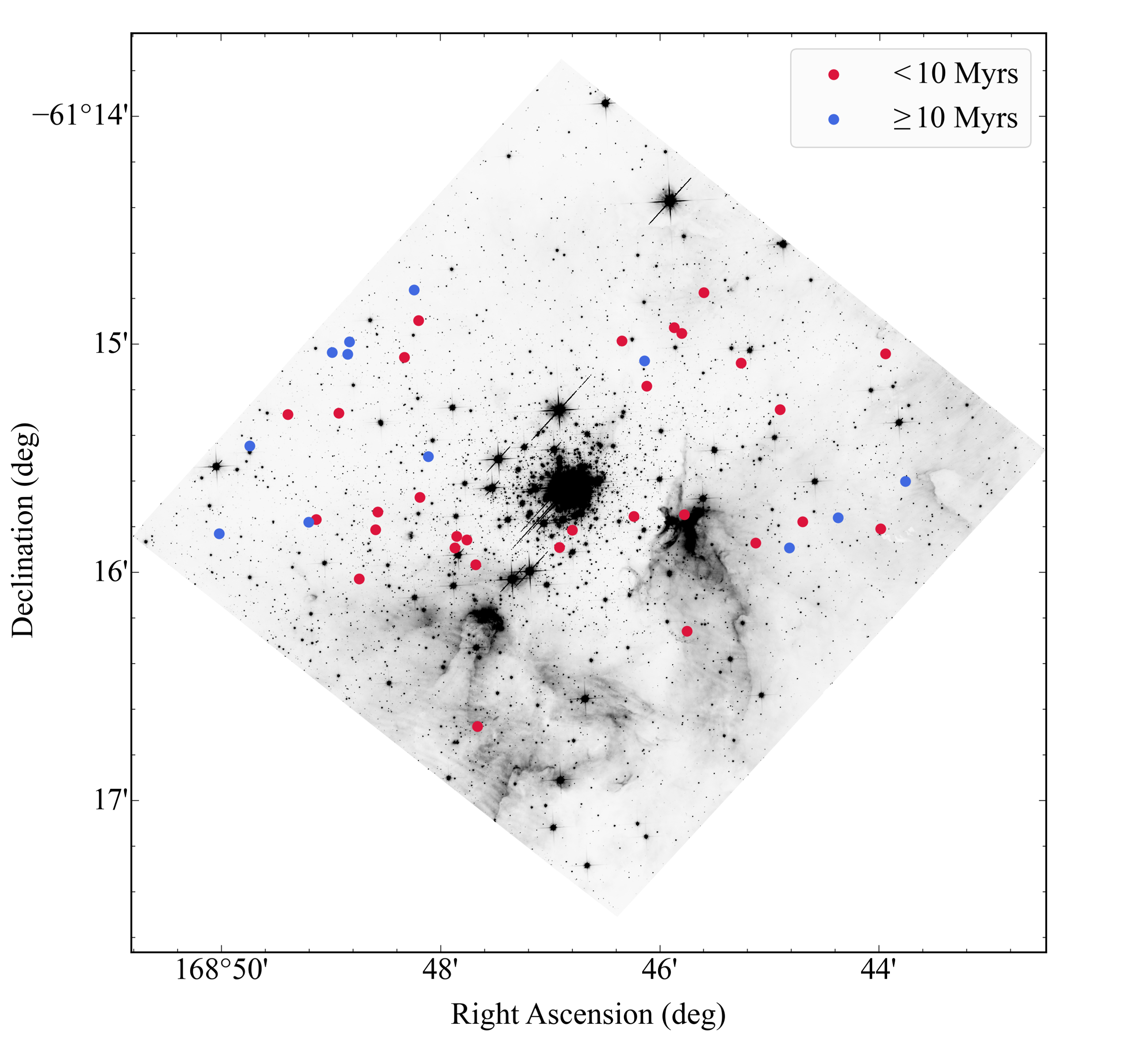}
    \caption{Location of accreting PMS stars in our sample. Sources with ages $<10$ Myr in red. Sources with ages $\ge10$ Myr in blue.}
    \label{fig:young_old_location}
\end{figure}

\begin{figure}[h]
    \centering
    \includegraphics[width=0.9\linewidth]{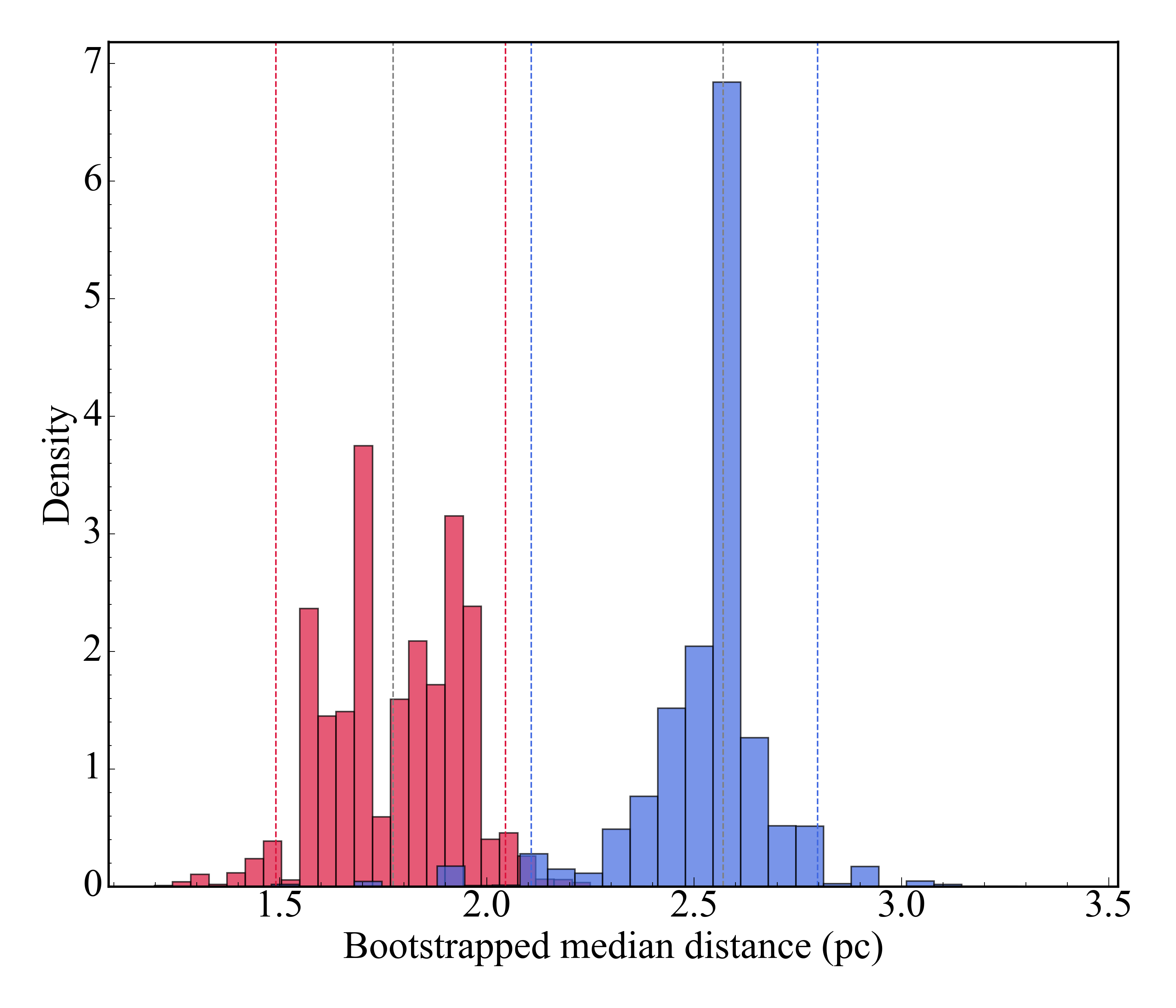}
    \caption{Results of bootstrapping analysis. The projected distances of young stars in red. Old stars in blue. $95\%$ confidence intervals shown as dashed red and blue lines. The median distance is shown as a dashed black line for each distribution.}
    \label{fig:bootstrap}
\end{figure}

We have plotted the location of the accreting PMS stars within NGC 3603 in figure \ref{fig:young_old_location}. Stars with ages of $<10$Myr, taking into account their uncertainties, are shown in red. The remaining older stars with ages of $\ge 10$Myr are shown in blue. The older stars tend to be farther away from the cluster centre, while the younger stars exhibit more scatter, but are generally closer to the centre.\\
In order to test whether there is a significant difference between the projected distances of the old and young PMS stars, we have performed two statistical tests. First, we performed a Kolmogorov--Smirnoff (K--S) test, which measures the maximum difference ($D$) between the cumulative distribution functions of two samples in order to investigate whether they come from the same underlying population. The test produces two results, the K--S statistic itself $D$, and a $p$-value, which indicates whether $D$ is the result of random chance. A $p$-value of $\le 0.05$ suggests that a significant difference exists between the two samples. We performed this test on the old and young stars by measuring their projected distances to the cluster centre. We obtained $D\;=\;0.550$, $p$-value$\;=\;0.007$. Although this result strongly suggests that the two samples are significantly different in terms of the distance to the cluster centre, our sample size is relatively small at just $42$. In an attempt to improve the statistical robustness of our small sample size, we employed bootstrapping to generate confidence intervals for the median distance to the cluster centre for the old and young stars. Bootstrapping involves resampling the projected distances with replacement many times, calculating a median distance each time. This generates a distribution of median distances for both the old and young stars. The two distributions are then compared. The degree to which the distributions overlap indicates whether they are significantly different from each other. The results from our bootstrapping analysis are shown in figure \ref{fig:bootstrap}. The younger stars have a median projected distance of $1.77$pc, with a $95\%$ confidence interval ranging from $1.49$ to $2.04$\,pc. The older stars have a median projected distance of $2.57$pc with a $95\%$ confidence interval ranging from $2.11$ to $2.80$\,pc. As these distributions do not overlap, this test along with the K--S test indicates that there is a meaningful difference in the projected distances of the old and young stars to the cluster centre.
\subsection{Mass accretion rates}
\label{subsec:acc_lum}
\begin{figure}
    \centering
    \includegraphics[width=0.95\linewidth]{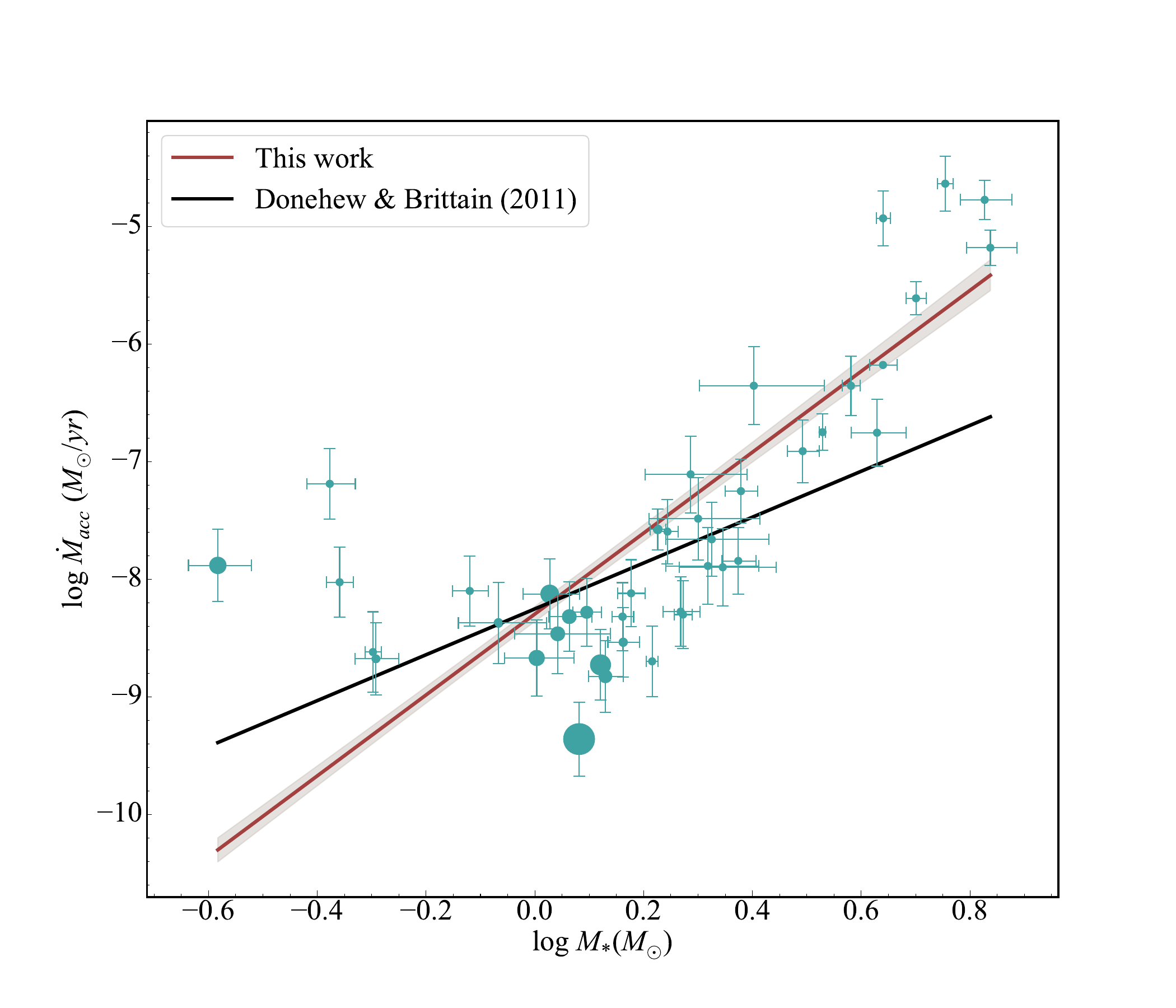}
    \caption{Accretion rate as a function of stellar mass. The size of each circle is proportional to age. The largest circle corresponds to $19$ Myrs. The best fitting line to our observations is shown in brown, while the best fitting line from D11 is shown in black.}
    \label{acc_versus_mass}
\end{figure}

\begin{figure}
    \centering
    \includegraphics[width=0.95\linewidth]{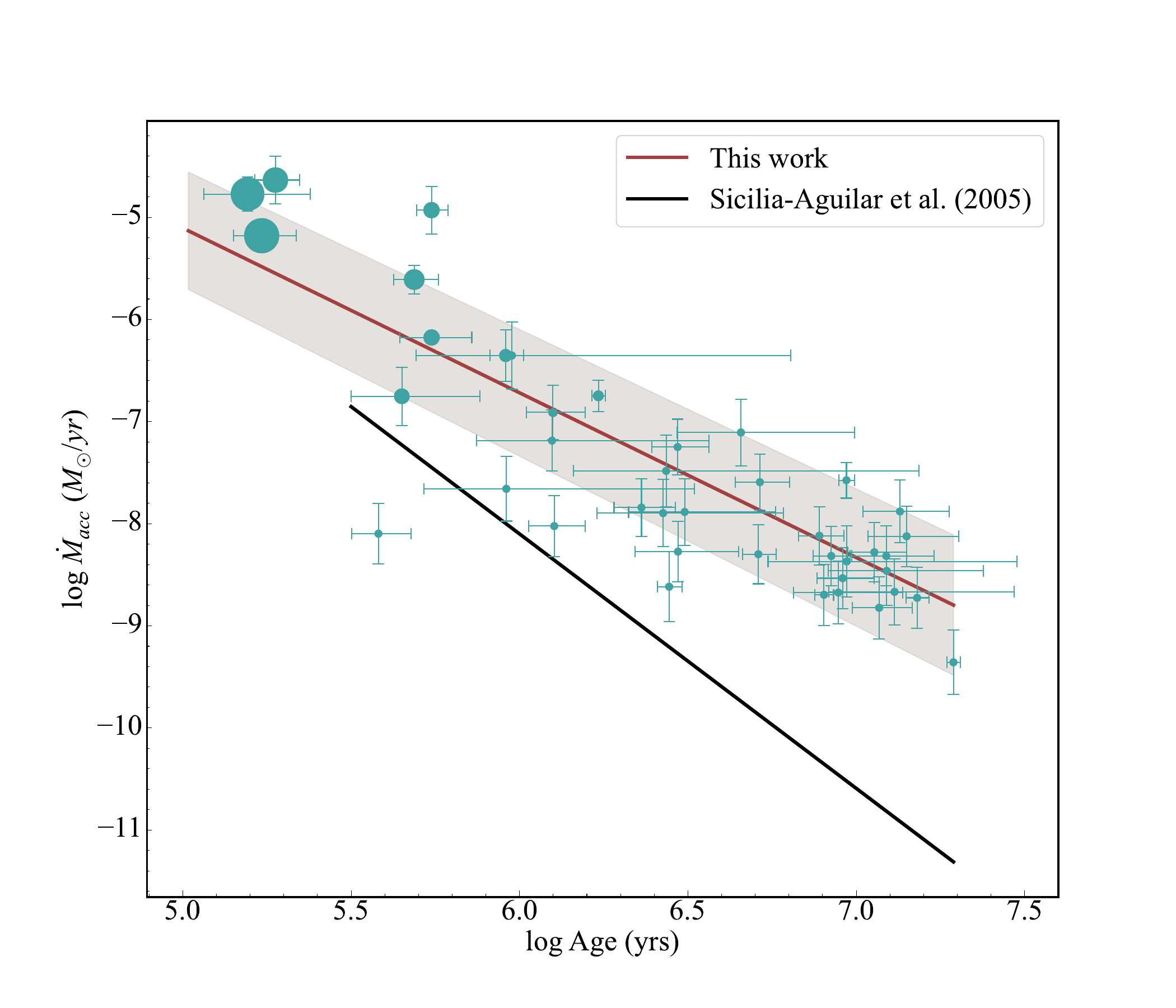}
    \caption{Accretion rate as a function of stellar age. The size of each circle is proportional to $M_{*}$. The largest circle corresponds to $7M_{\odot}$. The best fitting line to our observations is shown in brown. The best fitting line from \cite{sicilia2005cepheus} is shown in black.}
    \label{acc_versus_age}
\end{figure}

To compute the mass accretion rate $\dot{M}_{acc}$ of our sources, we first needed to calculate their accretion luminosity $L_{acc}$. To do this, we used the calibrations from \cite{alcala2017x} (A17) for CTTS and \cite{donehew2011measuring} (D11) for Herbig AeBe stars to convert the line luminosity of $Br_{7}$ to $L_{acc}$. Having determined the masses and radii of our sources, we converted $L_{acc}$ to $\dot{M}_{acc}$ using the formula,
\begin{equation} \label{eq_alcala}
    \centering
    \dot{M}_{acc} \sim 1.25 \times \frac{L_{acc}R_{*}}{GM_{*}} 
\end{equation}
from \cite{1998ApJ...492..323G, 1998ApJ...495..385H}. The emission line properties of our sources, including their $L_{acc}$ and resulting $\dot{M}_{acc}$ are given in table \ref{tab:emission_line_properties}. The values of $\dot{M}_{acc}$ span five orders of magnitude in our sample. Figure \ref{acc_versus_mass} shows $\dot{M}_{acc}$ of our sources with respect to $M_{*}$. Unsurprisingly, we see that $\dot{M}_{acc}$ increases with increasing $M_{*}$, though there is considerable scatter for the lowest and highest mass stars in our sample, well beyond their 1 $\sigma$ uncertainties. There is no apparent trend between $\dot{M}_{acc}$ and $M_{*}$ for the low mass end of our sample. For $M_{*} \ge 1 M_{\odot}$, the accretion rate increases sharply with increasing stellar mass, reaching a peak of $\dot{M}_{acc} = 10^{-4} M_{\odot} yr^{-1}$. We fitted a line to the relationship between $\dot{M}_{acc}$ and $M_{*}$, also shown in figure \ref{acc_versus_mass}. The best fitting coefficients to this line are
\begin{equation} \label{eq_alcala}
    \centering
    log \; \dot{M}_{acc} = 2.99 \; (\pm 0.112) \; log \; M_{\odot} \; -8.0673 \; (\pm 0.0525).
\end{equation}
We also include in figure \ref{acc_versus_mass} the relationship found between $\dot{M}_{acc}$ and $M_{*}$ from D11. In that study, the authors measured the accretion rates of Herbig AeBe stars, but included a figure which combined observations from Herbig AeBe stars, intermediate mass T Tauri stars \citep{2004AJ....128.1294C}, as well as CTTS \citep{muzerolle1998brgamma}, spanning a range of masses from $0.1-12\;M_{\odot}$, corresponding to more than 3 orders of magnitude in mass. These measurements come predominantly from nearby low-mass SFRs. The relationship found for our sample of stars in NGC 3603 is significantly steeper than that of D11. This indicates that, for a given mass, the stars in our sample are accreting at higher rates than those from the combined sample taken from D11.\\ 
In figure \ref{acc_versus_age} we show the relationship between $\dot{M}_{acc}$ with age. As expected, $\dot{M}_{acc}$ tends to decrease with age \citep[e.g.][]{alcala2014x, manara2015x, alcala2017x}. The recovery of this relationship demonstrates that residual nebular emission does not significantly affect our measurements. We have fitted the relationship between $\dot{M}_{acc}$ and age. The line of best fit is shown in figure \ref{acc_versus_age}. We have also included the line of best of fit reported in \cite{sicilia2005cepheus}, based on a combination of observations from several regions including Cepheus OB2, Taurus, Ophiuchus, Chamaeleon I, and TW Hydrae association \citep{muzerolle2000disk, 2000prpl.conf.....M}. We have only included the portion of this line that is linear in log-space, starting at $log\;age\;=\;5.5$. The $\dot{M}_{acc}$ versus age relationship found for our sample is offset and significantly less steep compared to the relationship reported in \cite{sicilia2005cepheus}.\\ 
We have demonstrated that our sample in NGC 3603 exhibits higher accretion rates for a given mass and age compared to nearby low-mass SFRs. We note that Cepheus OB2 is a massive SFR, harbouring the O6 star HD 206267. \cite{sicilia2005cepheus} draw attention to several sources from Cepheus OB2 which also exhibit high $\dot{M}_{acc}$ and old ages. This behaviour is not seen in the observations from low-mass SFRs.\\ 

\subsection{Environmental correlations with $\dot{M}_{acc}$}
\begin{figure}[!hbt]
    \centering
    \includegraphics[width=0.95\linewidth]{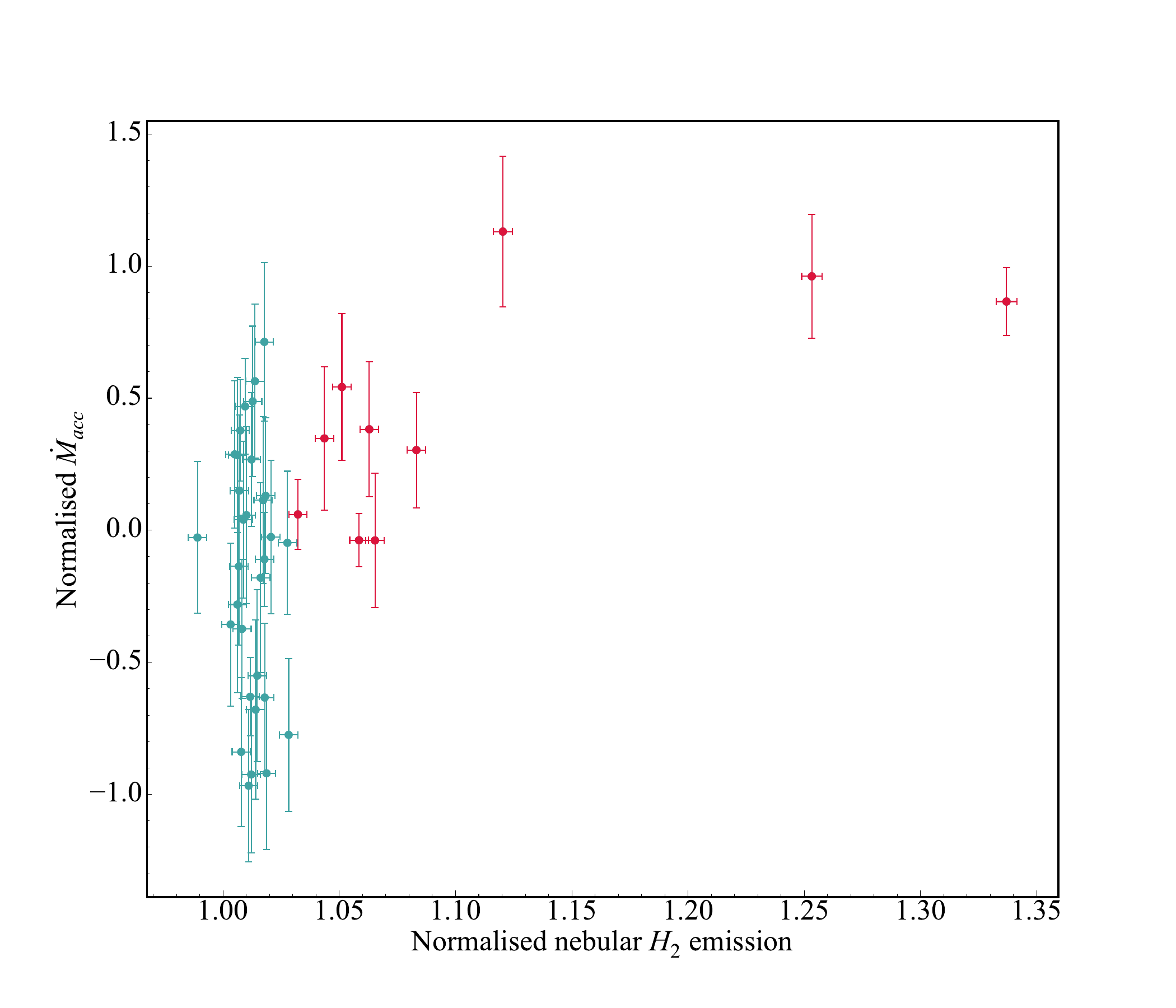}
    \caption{Normalised accretion rate as a function of the nebular $H_2$ brightness around each source. Sources associated with above average levels of nebular $H_2$ brightness are coloured red.}
    \label{acc_versus_H2}
\end{figure}

The mass accretion rates presented in D11 appear to follow a single relationship for the entire mass range $0.1 - 12$ $M_{\odot}$. Similar relationships have been found by \cite{muzerolle2003accretion, calvet2004mass, herczeg2008uv, fang2009star, alcala2014x, antoniucci2014poisson, manara2015x}, where a single power law can reasonably fit a wide range of stellar masses and accretion rates. There is significant scatter in the relationship that we have found between $M_{\odot}$ and $M_{*}$ for out sources. This motivated us to investigate whether there are environmental factors that correlate with the accretion rate, which may explain the scatter in figure \ref{acc_versus_mass}.\\ 
The environmental factors that we have considered are the ionised nebular gas emission, the molecular nebular gas emission, and the radial distance from the cluster centre. The ionised gas emission has been measured via the F656N $H_{\alpha}$ filter from HST WFC3. This was done by placing an annulus of $r_{in}=0.5\arcsec$, $r_{out}=0\farcs6$ around each source in the F656N image. These dimensions were chosen to ensure that light from the star had dropped off sufficiently with respect to the nebular emission. We calculated the median flux within this annulus, representing the typical nebular emission in $H_{\alpha}$ around each source. We have assumed that this emission is proportional to the density of the ionised nebular gas. The molecular gas emission was measured in the same way as $H_{\alpha}$, using a $H_{2}$ narrow band image centred at $2.12$ $\mu m$ from the High Acuity Wide field K-band Imager (HAWK-I) at the Very Large Telescope. In this case we set $r_{in}=2\arcsec$, $r_{out}=2.1\arcsec$, owing to the larger PSF of HAWK-I. Again, we have assumed that the $H_2$ emission is proportional to the molecular gas density. The radial distances were obtained simply using the Euclidean distance formula, using the F814W HST image, which highlights the continuum emission from the stars. \\
In order to investigate whether these external factors may influence accretion, we first needed to remove the influence of both $M_{*}$ and age. As pointed out by \cite{de2011photometric,de2017photometric}, this is necessary because the mass accretion rate depends simultaneously on both parameters. The sources in our sample cover a wide range of ages and a PMS star of a given mass will have progressively smaller mass accretion rate as it ages.  Following \cite{de2011photometric,de2017photometric}, we did this by performing a multivariate fit between $\dot{M}_{acc}$, $M_{*}$ and age $t_*$. The best fitting coefficients are,
\begin{equation} \label{multivariate_best_fit}
    \centering
    log \; \dot{M}_{acc} = 1.682 \; log \; M_{*} -0.795 \; log \; t_* -2.731
\end{equation}
We then calculated the residuals of $\log \; \dot{M}_{acc}$ and the best fitting line, and compared those residuals with the three environmental factors. We found no significant correlation between $\dot{M}_{acc}$ and ionised nebular gas emission, or the radial distance of the sources to the OB cluster. In the case of molecular gas emission however, we found a positive correlation with $\dot{M}_{acc}$, as shown in figure \ref{acc_versus_H2}. The majority of the sample are associated with comparable, low levels of $H_2$ emission, finding themselves in the low density bubble near the central cluster. These sources exhibit a wide range of $\dot{M}_{acc}$, indicating that a lack of high density molecular gas around a source does not imply a low accretion rate. However, in cases where accretion rates are higher than expected for a given age and mass, those same sources tend to be associated with regions of high density molecular gas, finding themselves in the prominent south-west and south-east pillars of the region.

\subsection{Revisiting the $L_{acc}$ -  $L_{Pa_{\alpha}}$ relationship}
\begin{figure*}[htbp]
    \centering
    \begin{minipage}[b]{0.45\textwidth}
        \centering
        \includegraphics[width=\textwidth]{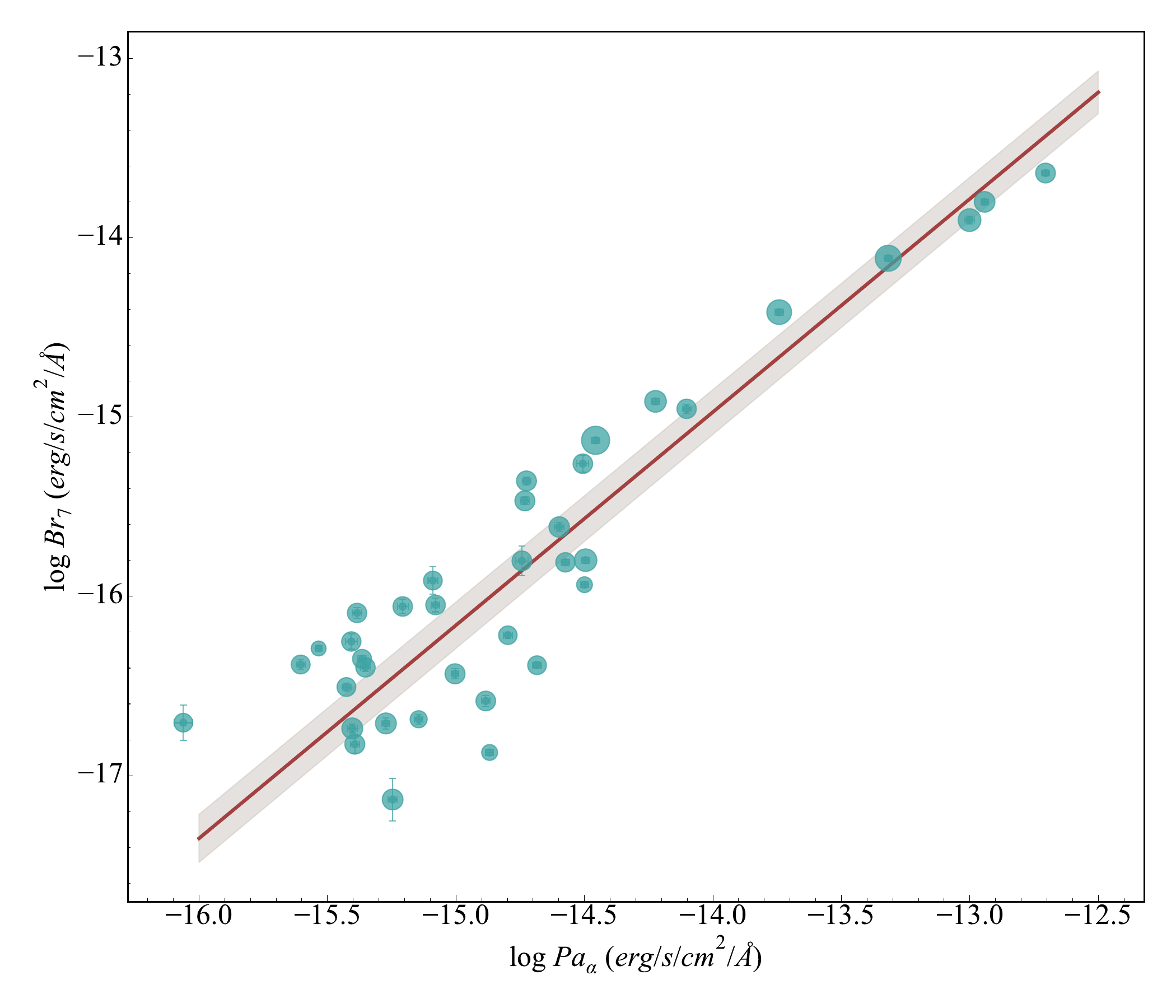}
        \caption{Relationship between the line flux of $Br_{7}$ and $Pa_{\alpha}$. The brown line represents the line of best fit and the grey shaded area shows the uncertainty of the fit.}
        \label{br_7_vs_pa_a}
    \end{minipage}
    \hfill
    \begin{minipage}[b]{0.45\textwidth}
        \centering
        \includegraphics[width=\linewidth]{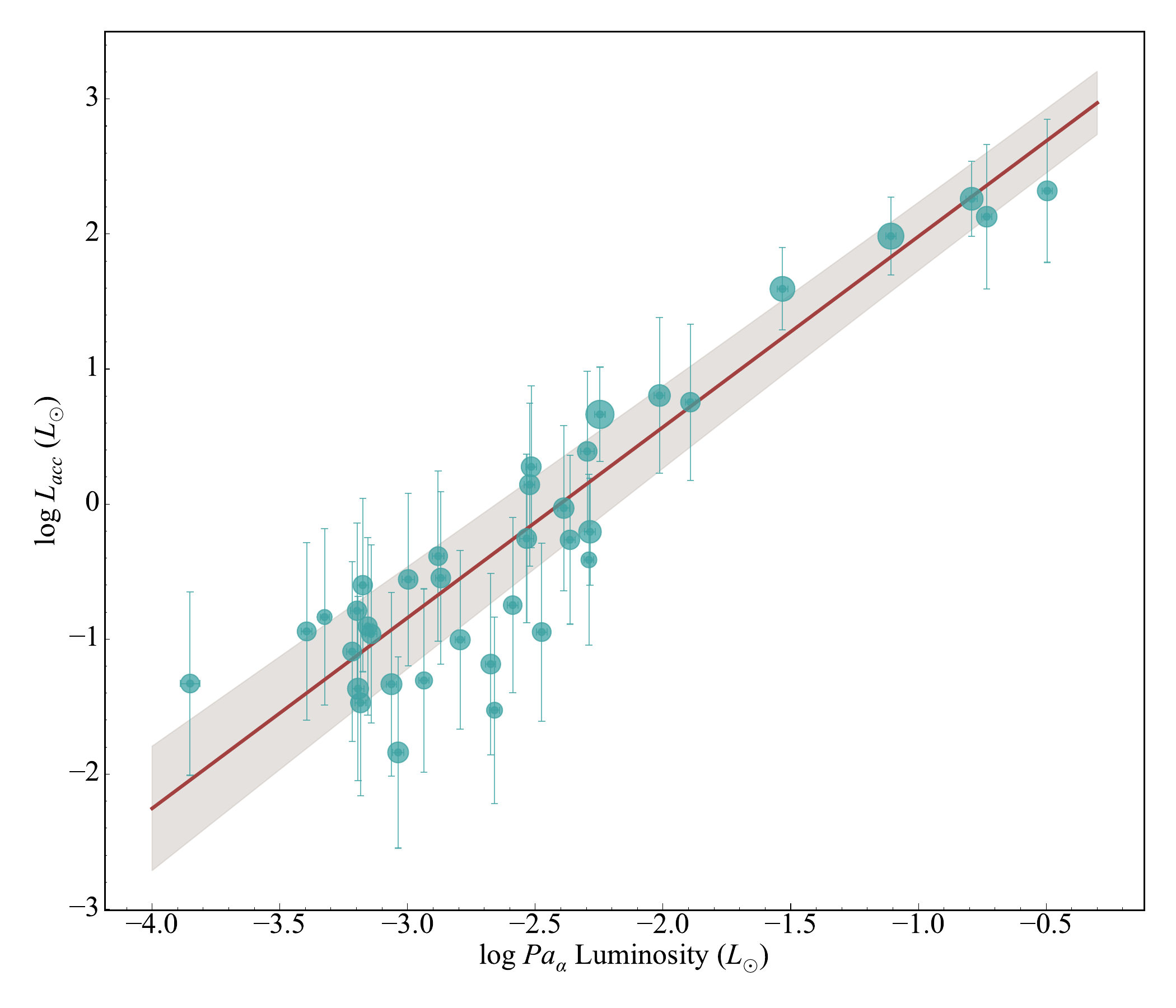}
        \caption{Relationship between $L_{acc}$ and $Pa_{\alpha}$. The brown line represents the line of best fit and the grey shaded area shows the uncertainty of the fit.}
        \label{acc_lum_vs_pa_a}
    \end{minipage}
\end{figure*}

In \cite{rogers2024determining}, we established the first relationship between $L_{acc}$ and $L_{Pa_{\alpha}}$. Since the completion of that work, we improved our methodology related to correcting for extinction, spectral classification, flux calibration and nebular subtraction. We present here an updated relationship. The net result of our changes and improvements in methodology is a significant reduction in the uncertainties of the best fitting line between $L_{acc}$ and $L_{Pa_{\alpha}}$, with no changes to the best fit coefficients.
\begin{equation} \label{revised_acc_lum}
    \centering
    log \; {L}_{acc} = 1.42 (\pm 0.08) \; log \; L_{Pa_{\alpha}} + 3.33 (\pm 0.17) 
\end{equation}
In figure \ref{br_7_vs_pa_a}, we show the relationship between the line fluxes for $Br_{7}$ and $Pa_{\alpha}$. Owing to our high quality observations and improved methodology we were able to measure both of these lines with high precision, leading to small statistical uncertainties for each flux measurement. To calculate $L_{acc}$, we converted our $Br_{7}$ line fluxes to luminosities, and then used the calibrations of A17 and D11 to convert to $L_{acc}$. The coefficients of the revised relationship are identical to \cite{rogers2024determining}, but with the uncertainties reduced by a factor $\sim$ 2.5. 
Figure \ref{acc_lum_vs_pa_a} shows the resulting relationship between $Pa_{\alpha}$ and $L_{acc}$. The best fitting line coefficients are given in equation \ref{revised_acc_lum}. What is clear by comparing figures \ref{br_7_vs_pa_a} and \ref{acc_lum_vs_pa_a} side by side is that the dominant uncertainty in this relationship is inherited from the conversion to $L_{acc}$ from A17 and D11. This is a result of the more challenging near-UV shock modelling approach that is used to directly determine $L_{acc}$ in A17 and D11. This approach unambiguously probes the material being accreted onto the star, but high extinction in the UV as well as absorption by Earth's atmosphere necessarily leads to larger statistical uncertainties. 

\section{Discussion} \label{sec:discussion}
In this section we discuss and interpret our results from above. This includes a brief review of the metallicity of NGC 3603 from previous studies, a discussion of disk tracing features detected in the NIRSpec spectra as well as the conspicuous lack of circumstellar $H_2$ emission. The impact of external photoevaporation on our sources is considered and the ages of our sources are discussed and compared to stars in low-mass SFRs. The possible implications of the environmental correlation between $\dot{M}_{acc}$ and nebular $H_2$ emission are considered.
\subsection{The metallicity of NGC 3603} \label{subsec:metallicity}
We opted to fit our source spectra with only solar metallicity models corresponding to [Fe/H] = 0.0. The metallicity of NGC 3603 has been investigated before, typically by obtaining nebular spectra of the H II region, and using recombination lines from metals like Ne and O to infer a metallicity \citep[e.g.][]{melnick1989galactic, lebouteiller2008chemical}. These studies report that NGC 3603 appears to be consistent with solar metallicity. As such, we have a strong prior belief that the metallicity of the stars should also be consistent with solar metallicity. PMS spectra are not well suited to a rigorous metallicity investigation, despite there being a number of strong metal absorption lines in our wavelength range from Al, Mg, Ca and Na. This is because the strength of metal absorption lines can be reduced due to NIR veiling from the protoplanetary disk. If the veiling is not fully characterised, it can lead to the appearance of a low metallicity due to apparently weak metal absorption. The metal species mentioned are most strongly in absorption for sources with low effective temperatures $3500 \le T_{eff} \le 5000$\,K. At higher temperatures, these metals can no longer survive in the stellar photosphere. For mid-G type stars and earlier at solar metallicity, metal lines are typically no longer visible \citep{husser2013new}. In the case of MS stars, it is therefore only the stellar temperature and metallicity that can affect the metal line absorption depths. Conversely, in the case of PMS stars, an additional variable is veiling, which itself is challenging to fully characterise. 
A possible solution to remove the influence of veiling in determining the metallicity is to look for metal lines, close to each other in wavelength, whose ratio changes with metallicity. As the lines are near each other, they experience approximately the same amount of veiling, and so their ratio should not change due to veiling. We did find one such ratio for the Al lines at $1.67189$ and $1.67505$ $\mu m$. Their ratio inverts when going from [Fe/H] = 0.0 to [Fe/H] = +1.0. However, none of our sources were well fitted with super-solar metallicity, and so this ratio was not used. We did attempt to fit a number of sources with sub-solar metallicity models, at $[Fe/H] = -1.0$, but the goodness of fit was comparable to the solar metallicity model best fit. Therefore, we have no reason to believe that the metallicity of our sources in NGC 3603 is different from the solar value.\\
\subsection{Hydrogen recombination lines}
\label{subsec:h_lines}
\begin{figure}[h]
    \centering
    \includegraphics[width=1\linewidth]{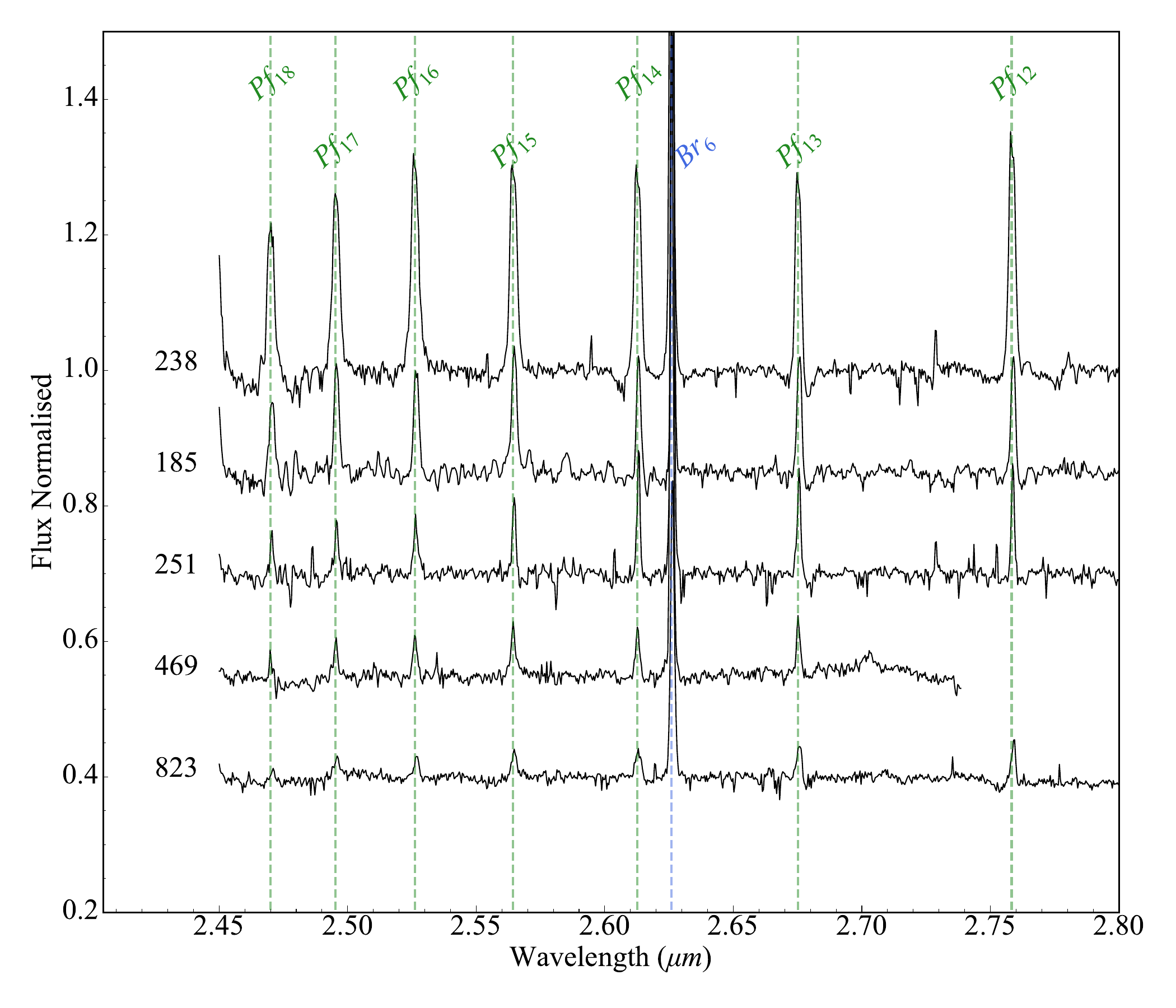}
    \caption{Five PMS sources that display Pfund recombination lines. Their spectra have been normalised to unity, and offset in this figure for visual clarity. The Pfund lines are marked with a green dashed line. $Br_6$ is also present, marked with a blue dashed line. Note that a number of pseudo absorption features are artificially created as a result of our normalisation.}
    \label{fig:pfund_lines}
\end{figure}

All 42 PMS spectra display some or all of $Pa_{\alpha}$, $Br_{6}$ and $Br_{7}$ in emission. $Pa_{\alpha}$ $\lambda\;1.875$ $\mu m$ and $Br_{6}$ $\lambda\;2.662$ $\mu m$ are two strong NIR lines that fall within wavelength ranges that are not typically recoverable from the ground due to telluric absorption. While it is possible to attempt to remove these telluric features with post-processing \citep[e.g.][]{vacca2003method, kausch2015molecfit}, the wavelength ranges of $1.8-1.9 \mu m$ and $2.5-2.7 \mu m$ are often omitted in NIR spectroscopic studies. JWST NIRSpec provides us with access to both of these lines simultaneously for the first time. This has facilitated the calibration of both of these lines as accretion tracers \cite{rogers2024determining}.
A subset of sources display additional Brackett lines as well as Pfund lines in emission. These sources are more massive and younger than the sample average. Both of these characteristics are typically accompanied by higher accretion and or ejection activity, which likely explains why these sources feature more line emission. The Pfund series is $\sim 3$ times weaker than the Brackett series, making it observationally challenging to detect with high S/N. Pfund lines have been detected with ground based observatories in star forming regions within $\le 1000$ pc \citep{maaskant2011sequential, koutoulaki2018circumstellar}, as well as in more distant HII regions from massive young stars \citep{bik2006vlt}, and have been detected in CTTS stars in the Small Magellanic Cloud \citet{jones2022near}. In these cases only a limited selection of the lines are visible, again due to telluric absorption. 
The uninterrupted Pfund series has been studied with space based telescopes such as the Infrared Space Observatory (ISO) \citet{benedettini1998iso}. Due to the relative weakness of the series as well telluric absorption, the Pfund series is not as well studied as other NIR or optical hydrogen series. In \cite{salyk2013measuring}, the authors showed that $Pf_{8}$ ($\lambda\;4.6538 \mu m$) can be used to measure the accretion rate of PMS stars, making it a useful diagnostic for deeply embedded objects as extinction at this wavelength is extremely low compared to optical wavelengths. In \cite{rogers2024kinematicevidencemagnetosphericaccretion}, the full width at half maximum (FWHM) of Pfund and Brackett emission lines was used to kinematically infer that these lines have an origin in the magnetospheric accretion flow, rather than from winds for a sample of intermediate mass stars, including a number of Herbig Ae stars. The Pfund series is shown in figure \ref{fig:pfund_lines} for a number of sources.
\subsection{Disk tracing species}
\subsubsection{CO bandheads}
\begin{figure}[h]
    \centering
    \includegraphics[width=1\linewidth]{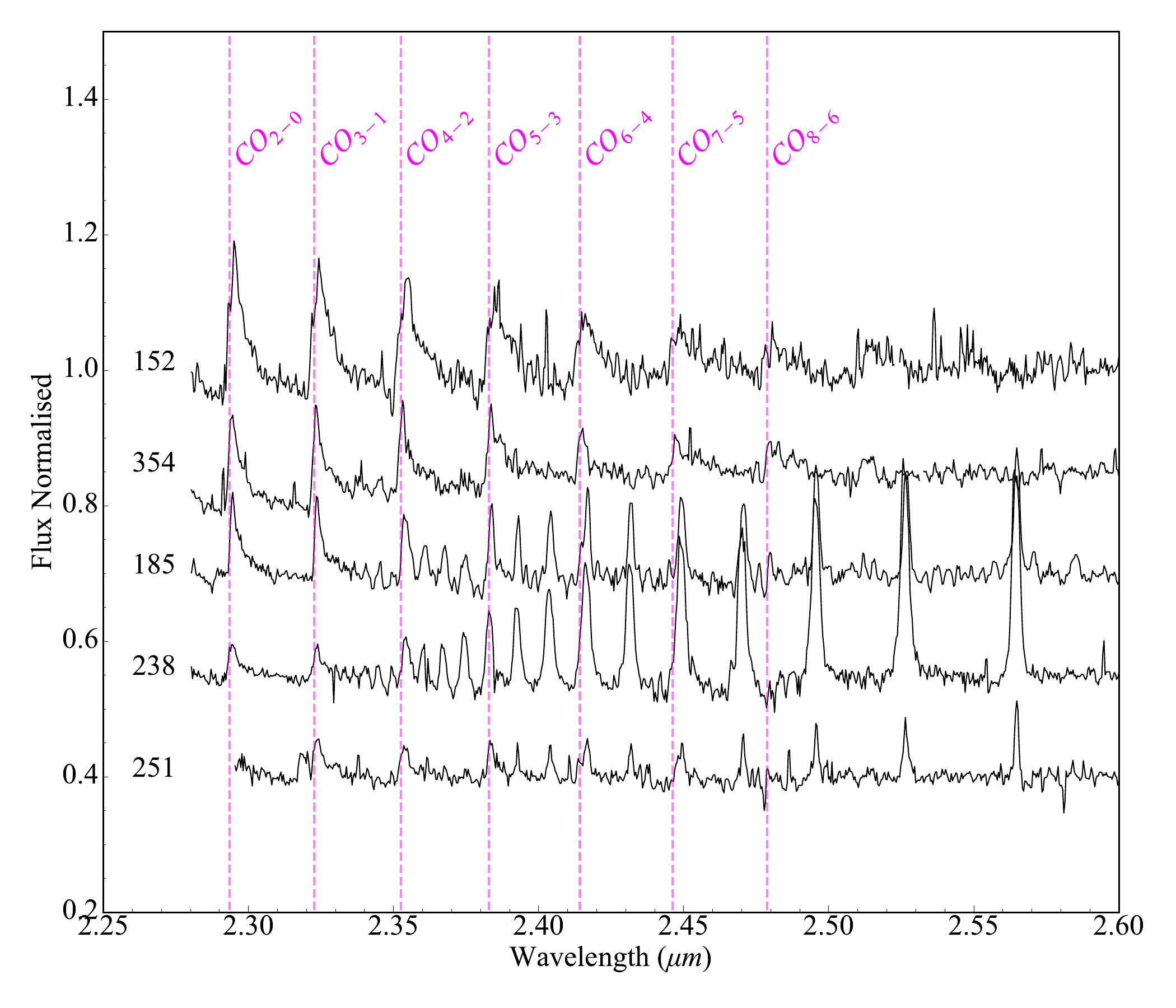}
    \caption{The normalised spectra of the five CO bandhead emission sources. The spectra have been offset for visual clarity.}
    \label{fig:CO_emission_norm}
\end{figure}

Disk tracing species Fe II $1.688$ $\mu m$ and CO bandheads are detected in emission in a number of our PMS spectra. CO bandhead emission is thought to arise from within the dust sublimation radius of the protoplanetary disk, at relatively high temperatures ($2000 - 5000$\,K) and densities ($n_H \ge 10^{10}$ cm$^{-3}$). An origin in the protoplanetary disk has been supported by studies at high spectral \citep{ilee2013co, bik2004evidence} and spatial \citep{o2020gravity, koutoulaki2021gravity} resolution. It is thought that self shielding occurs at these high densities, which protects the bulk of the CO from dissociation via UV photons from the central star (and in the case of our sources in NGC 3603, from the OB cluster). The profile of the CO bandheads is typically composed of a 'blue shoulder' due to doppler broadening of the line \citep{ilee2013co}. The red wing of the line is composed of a series of closely spaced vibrational transitions, known as J-lines. The J-lines are not resolved at NIRSpec's spectral resolution, giving the bandhead the appearance of a long red tail.\\
Of the 42 sources in this study, 5 show CO bandhead emission ($12\%)$. This is below the detection rate of $\sim 25\%$ found in \cite{ilee2013co}. These sources are $152$, $185$, $238$, $251$, and $354$. All of them are of intermediate mass, ranging from 4 to 7 $M_{\odot}$. For sources $152$ and $354$, the first 6 bandheads are clearly detected. For both of these sources, there is tentative evidence of the blue-shoulder, especially in the strongest transition $CO_{2-0}$. In the cases of sources $185$, $238$ and $251$, the bandheads are noticeably weaker in terms of their EW, and no blue shoulder is detected. The weaker appearance of their bandheads is likely a result the stronger stellar continuum, as these sources are twice as massive as $152$ and $354$. This leads to veiling of the bandheads, reducing their apparent strength. All sources with CO bandhead emission also display strong hydrogen recombination lines and exhibit significant NIR excess, to the extent that no absorption lines are detected in their spectra. Figure \ref{fig:CO_emission_norm} shows the rectified CO bandhead spectrum of each source, with an offset on the y-axis for clarity.\\
\subsubsection{Fe II}
The Fe II fluorescent emission line at $1.688$ $\mu m$ is detected in just three of our sources: $251$, $469$, and $823$. This line is more commonly detected in intermediate and high mass PMS stars \citep{porter1998broad}, as it is likely pumped by Lyman photons, and hence a strong UV source is required. This is true for our sample of stars, with the line being detected in some of our most massive sources. As Fe is easily ionised to high ionisation states, in order for Fe II to survive, it must reside in dense gas. It is thought to trace dense material of the inner disk, in a region outside of where the CO bandheads are produced \citep{lumsden2012tracers}, and is known to originate in the inner disk of classical Be stars \citep{carciofi2006non}. As such, its origin in the protoplanetary disk of PMS stars seems reasonable. This line is located directly beside $Br_{11}$, and is notably narrower than $Br_{11}$ in all three sources. This supports the idea that Fe II is not produced in an outflow or accretion flow, as the high velocities involved would cause considerable line broadening. The presence of CO bandheads and Fe II in conjunction with strong continuum NIR excess indicates that sources with sufficient mass can retain a significant gas rich inner disk while being strongly irradiated by massive stars for several $10^5$ years.
\subsection{Lack of circumstellar $H_2$}
Despite our large, high S/N sample of PMS sources of varying masses and stages of evolution, we did not detect any $H_2$ emission from any of our sources. The following discussion is largely qualitative in nature, and is not intended as a robust explanation to the lack of circumstellar $H_2$, but rather to consider some potentially important physical mechanisms that could be at play for our sources.\\ 
$H_2$ is commonly seen in emission from PMS stars, coming either from the protoplanetary disk itself \citep{weintraub2000detection, bary2002detection} or, more commonly, tracing a wind or outflow \citep{takami2007micro, beck2008spatially, harsono2023jwst}. We detect numerous $H_2$ emission lines in the nebular spectra. The nebular spectra that show these lines most prominently are associated with the dense pillars of material seen in the south-west and south-east of NGC 3603. However, the absence of circumstellar $H_2$ could be due to a number of reasons. \\
Given the large distance to NGC 3603, emission from the protoplanetary disk or possible outflows is spatially unresolved out to distances of $\sim 700$ AU. Emission from the star, disk and possible outflow are therefore all contained within the same beam. If the emission from $H_2$ is intrinsically faint, it may be too weak to detect above the continuum level of the stars. On the other hand, the robust detection of CO bandheads, as well as other disk-tracing species such as Fe II $1.688$ $\mu m$ likely rules this explanation out. When these features are detected in the spectra of young stars, $H_2$ tends to also be present, and is generally as strong or stronger than the CO or Fe II \cite[e.g.][]{davis2011vlt}.\\ 
Another possible explanation is that, if the bulk of the $H_2$ emission would come from a wind or jet as is typically found for CTTS and Herbig AeBe stars, this outflow would experience external photoevaporation from the central OB stars. This could dissociate or even ionise the outflows. Externally ionised Herbig-Haro jets have been observed from numerous young sources in Orion \citep{reipurth1998protostellar, bally2001irradiated, bally2003irradiated}. Those observations spatially resolved the jet from the central star, and unlike typical molecular Herbig-Haro outflows, the spectra of those jets resemble nebular emission spectra from photoionised gas. It is therefore likely that a similar, or possibly even enhanced form of this jet ionisation occurs also in NGC 3603, given its much larger population of early O type and Wolf-Rayet stars compared to Orion. In \cite{reipurth1998protostellar}, the authors even suggest that launching of the jet may be entirely quenched on the side that faces the OB cluster, as the four sources in their study show significantly less developed jet lobes on the directly irradiated side, while the more prominent lobes face away, possibly shadowed by the protoplanetary disk itself. \\
The recent work of \cite{kirwan2023spectacular} partially answers this question some twenty years later. Here, the authors studied the enigmatic Orion proplyd $244–440$, which launches a prominent bipolar jet that is apparently unaffected by the UV environment. They reason that the jet was likely launched after the onset of external photoevaporation, and is shielded from much of the UV radiation by the envelope of the proplyd itself. This protection can only last until the jet reaches the ionisation front and penetrates the cusp of the proplyd envelope, after which it is exposed to the external environment. It seems plausible then that in NGC 3603 any $H_2$ outflows could have already been destroyed given the four orders of magnitude more intense UV field compared to Orion \citep{rollig2011photon}, which would result in an ionisation front closer to the disk surface \cite{winter2022external}. \\
This is consistent with the work of \cite{reiter2013hst} and \cite{reiter2016fe}, who used HST narrow band imaging of $H_{\alpha}$ and [Fe II] $1.644$ $\mu m$ (not to be confused with the permitted Fe II $1.688$ $\mu m$) to detect and characterise externally ionised outflows from young stars in the Carina star forming association near the massive SFR Trumpler 14. Trumpler 14 contains at least 74 O stars \citep{berlanas2023gaia}. The existence of prominent, highly collimated jets in this region demonstrates that jet launching need not be entirely switched off in the presence of high UV radiation. These jets are consistent with the externally ionised jets found by\cite{reipurth1998protostellar}, producing predominantly atomic and ionic emission lines. In a more recent work using the JWST Early Release Observations (ERO) of NGC 3324, adjacent to the Carina nebula, \cite{reiter2022deep} detect outflows traced by both $Pa_{\alpha}$ as well as $H_2$, with the vast majority of the $H_2$ outflows being found within molecular clouds where they are presumably protected from the harsh UV environment.\\
In summary, we suggest that the lack of circumstellar $H_2$ detected in our sample may result from external ionisation of any initially molecular outflows by the central OB cluster, as has been seen in other massive star forming regions.
\subsection{External photoevaporation in NGC 3603}
\label{subsec:ext_photoevap}
External photoevaporation is the defining process associated with the objects known as proplyds. Proplyds are characterised by their unique teardrop shaped cocoon, enveloping the protoplanetary disk and star, formed from material boiled off of the disk surface. For NGC 3603, we lack the spatial resolution to classify our sources as proplyds based on this criterium (although based on the total lack of circumstellar $H_2$ emission, it is highly unlikely that our sources are similar to the proplyds found throughout Orion). Regardless, our sources can unambiguously be classified as externally irradiated stars. As such, external photoevaporation has likely played a role in the evolution of these sources at some point in their lifetime. \\
External photoevaporation is a possible explanation for the low levels of NIR excess that we have measured towards the majority of our sources. This process truncates the protoplanetary disk, eroding the disk from the outside in \citep[e.g.][]{ndugu2024external}. A smaller disk results in a smaller emitting area, and hence less NIR excess emission. Disk sizes in the Orion Nebular Cluster have been shown to be smaller than in other SFRs and even to scale with the distance to the central ionising OB stars \citep{boyden2020protoplanetary}. With that in mind, most of the NIR excess is thought to originate from the inner disk within $\sim$1 AU of the star. Material this close to the star is located deep in the potential well, and is therefore more difficult to photoevaporate. The disks in NGC 3603 are more strongly irradiated compared to the disks in Orion. They will therefore have higher sound speeds due to higher temperatures. This serves to reduce the gravitational radius \citep{winter2022external}, which could allow for inner disk depletion. It has also been shown in \cite{haworth2018fried} that in high $G_{0}$ environments, even the inner disk can be significantly impacted by external photoevaporation.\\
Without exact knowledge of the UV field at the surface of the disk, it is unclear whether a significant fraction of the inner disk could be lost due to external photoevaporation, and whether this could explain the lack of excess emission at 2-3 $\mu m$. Several of the results that we have presented here also seem to contradict significantly depleted disks. The disk lifetime is expected to be dramatically reduced due to external photoevaporation \citep{henney1999keck}, but we have determined that many of our sources are still accreting after 10 or even 20 Myr. Additionally, the accretion rates of our sources are higher than stars with comparable masses and ages based on other studies in the literature. This has already been been pointed out by \cite{de2017photometric} from the $H\alpha$ excess of sources in other massive star forming regions. Both of these results suggest that a robust gaseous disk survives around our sources. As such, it remains unclear how external photoevaporation has impacted our sources. Observations at longer wavelengths, for example with JWST's MIRI, would directly probe more complex chemical species in the disks. These molecular features are expected to change significantly as a result of UV driven photochemistry \citep{walsh2012chemical, walsh2013molecular}, though recent results from MIRI seem to contradict this \cite{ramirez2023xue}. Follow up observations at longer wavelengths could also reveal to what extent the disks may have been truncated due to external photoevaporation.

\subsection{How long can accretion last for stars in NGC 3603?}
In low-mass SFRs with low levels of UV radiation, disk dispersal timescales are typically around $5$ to $10$ Myr \cite[e.g.][]{armitage2003dispersion}. Long lived disks with ages of $40$ to $50$ Myr have been found around very low mass stars, so called ``Peter-Pan'' stars. This has been attributed to below average levels of UV radiation in their environment \citep{coleman2020peter}, resulting in lower than average levels of photoevaporation. It is therefore surprising that protoplanetary disks undergoing active accretion could survive for $\sim 15$ Myr in an environment like NGC 3603, which has some of the highest levels of UV radiation in the Galaxy \citep{melena2008massive}. The disk fraction in such an environment is expected to be close to $0\%$ after $7$ Myr \citep{guarcello2010chronology}.\\
This result of long lived disks in NGC 3603 is in agreement with the photometric study of B10, who found evidence of an old PMS population still undergoing accretion based on $H_{\alpha}$ excess emission using HST narrow band imaging. In that study, a single average extinction correction was applied to all sources of $A(V)$ = 4.5, using the extinction curve of \cite{cardelli1989relationship} with $R(V)$ = 3.55. We have determined the extinction spectroscopically for each source individually, and have used a higher value of $R(V) = 4.8$. The average value of A(V) that we found was $4.04$, with a dispersion of $2.1$ across the sources. The lower value of $R(V)$ and higher value of $A(V)$ used in B10 tends to push the sources towards bluer colours, making them appear older and closer to the zero-age-main-sequence (ZAMS) line. Indeed, if we apply the extinction correction to our sources as was done in B10, our old sources move from ages of $\sim 15$ Myr to $\sim 50$ Myr. Despite these differences in methodology, our spectroscopic findings are in general agreement with B10, that accreting PMS stars survive in NGC 3603 well beyond the expected cut-off point of $\sim 7$ Myr.\\
Disk dispersal timescales have been inferred observationally by searching for stars with NIR excess emission in star forming regions with different ages \citep{strom1989circumstellar, haisch2001disk, hillenbrand2005observational, yasui2009lifetime}. Searching for NIR excess could be done photometrically with $J,H,K$ bands, making it an efficient method to probe disk lifetimes in many different star forming regions. We have found spectroscopically that 26 of our 42 PMS stars show no detectable NIR excess in the $J,H,K$ bands, but do show strong hydrogen recombination lines indicating ongoing accretion. We argue that emission lines provide a more sensitive means of inferring ongoing accretion, as these lines more directly probe accretion related activity compared to NIR excess, and are easier to detect in spectra compared to weak excess emission associated with the later stages of star formation.\\
\cite{thanathibodee2018evolution} studied three sources with low mass accretion rates from the UV to NIR. Two of the sources showed unambiguous evidence of ongoing, albeit weak, accretion based not only on hydrogen emission lines, but also redshifted absorption in the NIR He I $\lambda 1.083$ $\mu m$ line, tracing infalling material. These sources showed negligible NIR excess, and a modest reservoir of remaining gas in the inner disk based on marginal detections of FUV $H_2$ emission. More sensitive facilities like JWST can detect faint spectroscopic signatures of accretion that were not captured by earlier efforts using broadband NIR excess.\\
A recent paper from \cite{winter2024spatially} (W24) has examined whether PMS stars in regions of enhanced interstellar gas and dust density could experience disk replenishment by infall from the interstellar medium. They found a statistically significant positive correlation between the mass accretion rate of the sources, and the gas density around those sources. This result demonstrated that a spatial correlation exists with the accretion rate of stars. Infalling streams of cold molecular material have been detected towards numerous PMS stars \citep[e.g.][]{tobin2024observational}. Their detection suggests that disk replenishment is a feasible mechanism to extend the lifetime of protoplanetary disks. Late stage infall could help explain why protoplanetary disks in massive SFRs can survive for longer than expected, even under the influence of external photoevaporation.\\
In a recent study, \cite{demarchi2024} presented a sample of old ($\sim$ $30$ Myr) accreting PMS stars in the massive SFR NGC 346 in the Small Magellanic Cloud using NIRSpec MSA observations. That study employed a similar approach to determining the age of their stars as we have done in this study. Those authors have studied the Magellanic Clouds extensively via HST photometry \citep{sabbi2006past, de2011photometric, de2013photometric, de2017photometric, biazzo2019photometric, tsilia2023photometric}, and have consistently found populations of accreting stars with ages well beyond what is thought possible based on current disk evolution theory. We note for the reader that many of these same authors conducted the survey presented in B10. The observations that we have presented here as well as those presented in \cite{demarchi2024} serve as spectroscopic follow ups to those photometric studies that claimed to have found old, strongly accreting PMS stars based purely on photometric measurements. These follow up observations prove that the results inferred from photometry were largely accurate, that disks around PMS stars can survive for much longer than we think, and indicate that there is are significant gaps in our understanding about how stars and protoplanetary disks form and evolve in massive SFRs.\\
Combining the high sensitivity of NIRSpec with the use of emission lines as disk tracers has enabled the detection of accreting PMS stars that would have largely gone undetected based solely on NIR continuum excess. Disk fractions inferred purely from NIR continuum excess are likely underestimates. We have shown that most of our sources show no detectable excess in the $J,H,K$ bands, but still have high accretion rates.

\subsection{How accurate are stellar isochrones?}
The use of stellar isochrones and evolutionary tracks are commonly employed in star formation studies to determine stellar properties like age and mass. Mass determinations from PMS evolutionary tracks are relatively reliable, with accuracies of $\sim$ $10\%$ \citep{stassun2014empirical}. Age estimates on the other hand typically carry larger uncertainties of a factor 2-3 \citep{soderblom2010ages}. In \cite{bell2013pre}, the authors re-assessed the ages of 13 star forming regions by fitting the MS population with stellar evolutionary models, and fitting the PMS population with their own semi-empirical PMS isochrones, finding good agreement between the two methods, which each rely on different stellar physics. The ages determined by the authors were typically $\sim$ 2 times older than previous estimates in the literature, indicating that stellar isochrones tend to underestimate the true age of PMS stars. If this can be applied to the MIST isochrones that we have employed, then our stars with ages of $\sim$ 15 Myr will in fact be closer to $\sim$ 30 Myr in age. Indeed, the point is raised by \cite{bell2013pre} that the timescale of star formation may be longer than we think due to a systematic error in isochronal ages. This systematic error serves to push stars towards younger ages. This doesn't change the fact that the timescale for disk dispersal appears relatively consistent across numerous different isochronal models for many star formation regions, even if this agreement is only consistent on a relative level. Studies have found that after $\sim$ 5 Myr, only $20 \%$ of stars are still surrounded by a protoplanetary disk \citep[e.g.][]{carpenter2006evidence, dahm2007spitzer}, with this fraction being apparently even lower in massive SFRs \citep{guarcello2021dispersal}. As pointed out by \cite{demarchi2024}, stars with active accretion signatures in Galactic low-mass SFRs are not found in the region of the HRD where our oldest PMS stars are located.\\
While the ages of stars as determined by isochronal fitting may be systematically offset and inaccurate on an absolute level, the relative difference in ages demonstrates that the timescale of protoplanetary disk dispersal differs between stars in low-mass SFRs and massive SFRs.

\subsection{Systematic uncertainties affecting the HRD}
Systematic uncertainties also affect the position of stars in the HRD, such as uncertainty in the distance to the region as well as the intrinsic spread of distances to sources within the cluster. In a review of Scorpius-Centaurus, the nearest OB association \citep{preibisch2008nearest}, the authors point out that the physical spread of distances in Upper Scorpius is $\pm\;20$ pc, which is a relatively large fraction of the total distance to the region at $145$pc. This introduces a significant degree of uncertainty to the positions of the sources on the HRD. In regards to NGC 3603, this is less significant for two reasons. First, the distance to NGC 3603 is large, meaning that the intra-cluster distances would also need to be large in order for it to become an important source of uncertainty. Second, while Upper-Scorpius is a large association, NGC 3603 is a dense, compact cluster, with a cluster radius of $\sim5$ pc \citep{nurnberger2002infrared}. Based on this, the small differences in distance to the individual sources in our sample should not significantly affect their position on the HRD. A consideration that has not been discussed so far is multiplicity. It is likely that more than half of our sample are in binaries or higher order systems \citep{reipurth1993visual}. The presence of companions serves to increase the total luminosity of our sources. By not taking this into account in our analysis, we have undoubtedly introduced additional uncertainty into their positions on the HRD. However, given that this effect serves to push our sources up, away from the ZAMS, the age estimates are likely underestimated, rather than overestimated.
\subsection{Spatial distribution of PMS stars in NGC 3603} \label{subsec:spatial_dist}
A significant difference at the 2$\sigma$ level in the projected distances to the cluster centre has been found between our old and young PMS stars. The younger stars are more concentrated towards the centre of the cluster, while the older stars are typically farther away. This confirms the result from B10, who also reported a difference in the spatial distribution of young and old PMS stars in NGC 3603. Similar bimodal distributions have been found in other studies of PMS stars in massive SFRs \citep{de2011clues, gouliermis2012clustered, demarchi2024}.\\ 
This may simply be a result of older PMS stars having more time to move through the cluster compared to younger PMS stars. The bimodal distribution provides encouraging evidence that there is a meaningful difference between the two samples of PMS stars, and that our age determinations are robust. Our observations are intrinsically biased to not observe stars close to the cluster centre, as the crowding is too intense even for JWST. As such, we may miss older PMS stars located in the cluster centre. Of course, this bias also applies to the younger PMS stars, which may serve to cancel out any relative shift between the two distributions. 
\subsection{$\dot{M}_{acc}$ versus nebular $H_2$ emission} \label{subsec:acc_vs_H2}
We have found a relationship between $\dot{M}_{acc}$ and nebular $H_2$ emission. Sources that show enhanced $\dot{M}_{acc}$ relative to $M_{*}$ and age are usually associated with higher than average levels of nebular $H_2$ emission, suggesting an environmental influence on $\dot{M}_{acc}$. This relationship could be a result of high interstellar gas density that facilitates late-stage infall onto the protoplanetary disk, as has been seen in recent studies \citep[e.g.][]{cacciapuoti2024dusty}. This infall could replenish the disk, allowing accretion to occur at high levels for longer. Additionally, the higher gas densities may help to protect the disks from external photoevaporation, thus retaining a larger fraction of primordial disk material to accrete.\\ 
It is also possible that the relationship works in the opposite direction. Rather than enhanced levels of $\dot{M}_{acc}$ being a result of higher molecular gas densities (inferred from $H_2$ emission), the $H_2$ emission could instead be a result of enhanced levels of turbulence, X-ray and UV emission due to accretion and ejection processes from the stars. However, if the enhanced nebular $H_2$ emission were actually a result of high accretion rates, we would expect to see a similar trend in the nebular $H_{\alpha}$ emission, but this is not the case.\\ 
In W24 the authors found a significant correlation with the mass-normalised accretion rate and gas density traced by 100 $\mu m$ emission. We believe that we have discovered a similar relationship for the young stars in NGC 3603. These early results both indicate that the environment in which a star and protoplanetary disk form can significantly influence their evolution. While external photoevaporation is being considered more and more as an important environmental factor related to the formation and evolution of protoplanetary disks and planets themselves \citep[e.g.][]{somigliana2020effects, winter2022external, mauco2023testing, aru2024kaleidoscope}, the role of enhanced gas and dust densities associated with massive SFRs has received less attention. External photoevaporation serves to dramatically reduce the lifetime of protoplanetary disks, becoming the dominant disk dispersal mechanism even for modest UV radiation fields \citep[e.g.][]{miotello2012evidence}. A natural explanation as to how protoplanetary disks survive for a long time in high UV environments is from protection and/or replenishment from the interstellar medium, which is systematically more dense in massive SFRs \citep[e.g.][]{heiderman2010star}. Regions in which external photoevaporation is expected to be important will therefore also be associated with a higher density interstellar medium. These two properties may serve to offset each other, or as our results suggest, the high gas densities appear to dominate, allowing protoplanetary disks to survive for $\ge10$ of Myr. The relationship that we have discovered between nebular $H_2$ emission and $\dot{M}_{acc}$ serves as further motivation to understand how these various environmental factors ultimately influence the birth and evolution of stars and planets.

\section{Conclusions} \label{sec:conclusions}
We have presented our analysis of 42 PMS stars and their protoplanetary disks in NGC 3603 with JWST NIRSpec spectroscopy. Our main conclusions are the following:
\begin{enumerate}
  \item The nebular contribution to our stellar spectra has been removed using a scaled subtraction technique that accounts for spatial variations in nebulosity between adjacent micro shutters. 
  \item 42/100 stellar spectra exhibit strong hydrogen recombination lines after nebular subtraction.
  \item All PMS sources have been spectrally classified for the first time. The majority of CTTS have spectral types M, K, and G. The remaining are intermediate-mass stars, including a number of Herbig AeBe type stars.
  \item Emission from Fe II and CO bandheads are detected towards a number of young, intermediate mass sources. The presence of these features indicates that, even under the influence of external irradiation, a significant gas-rich inner disk can remain around sources with sufficient mass.
  \item No circumstellar $H_2$ emission is detected towards any source. The lack of this relatively common feature in CTTS and Herbig AeBe spectra may be the result of external irradiation of the outflow that $H_2$ typically traces.
  \item 26/42 sources have Class III SEDs based on their spectral index $\alpha$. 22 of those source spectra were best fitted with zero veiling emission added to their spectrum. The lack of excess emission indicates that these sources are at late stages of evolution, or their disks have been significantly truncated. 
  \item The ages, masses, and radii of all 42 sources have been determined with stellar isochrones and evolutionary tracks.
  \item The majority of sources have ages $\le 5$ Myr. 12/42 stars have ages 10-20 Myr. This is significantly longer than the expected disk dispersal timescale of $\le$ 7 Myr.
  \item The mass accretion rate of our sources spans five orders of magnitude. There is a strong relationship between $\dot{M}_{acc}$ and $M_{*}$ for stars $\ge$ 1 $M_{\odot}$. This relationship is significantly steeper than found in previous studies.
  \item $\dot{M}_{acc}$ is greater for stars in NGC 3603 for a given age and decreases with age more slowly compared to low-mass SFRs.
  \item The old PMS stars ($>10$ Myrs) appear significantly further away from the cluster centre compared to the young PMS stars.
  \item After removing the influence of both age and mass on $\dot{M}_{acc}$, we have found a relationship whereby sources associated with high density molecular gas tend to show enhanced $\dot{M}_{acc}$. This apparent environmental influence on $\dot{M}_{acc}$ may help to explain the high accretion rates and old ages of many of our sources.
\end{enumerate}

\section{Data Availability}
\href{https://zenodo.org/records/15188940}{Best fitting model spectra available on Zenodo.}

\begin{acknowledgements}
      We wish to thank the referee for their helpful comments which have improved the quality of this work.
{\em Software}: 
\texttt{SciPy} \citep{virtanen2020scipy};
\texttt{emcee} \citep{foreman2013emcee};
\texttt{NumPy} \citep{harris2020array};
\texttt{Matplotlib} \citep{hunter2007matplotlib}; 

\end{acknowledgements}

\bibliographystyle{aa}
\bibliography{references.bib}

\begin{appendix}
\section{Spectrum averaging} \label{subsec:app_spec_avg}
\begin{figure}[h]
    \centering
    \includegraphics[width=1\linewidth]{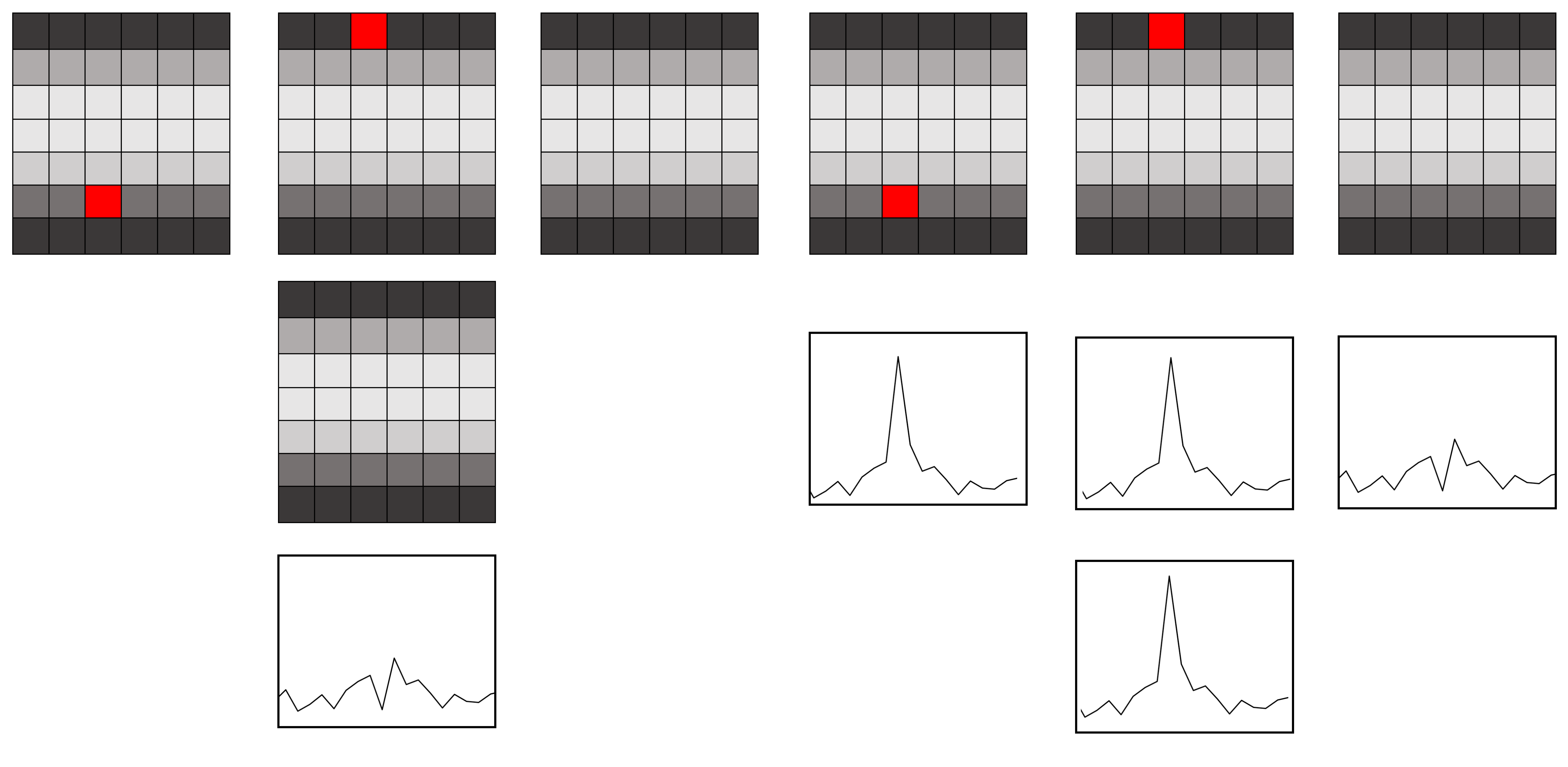}
    \caption{Left panel, (\textit{top}) shows three 2D spectra. Two contain a spike due to spurious pixels. (\textit{Middle}) resulting 2D spectrum after averaging. Spikes are no longer present, having been averaged out. (\textit{Bottom}) final median 1D spectrum with no spike. Right panel, (\textit{top}) shows three 2D spectra. Two contain a spike due to a spurious pixel. (\textit{Middle}) resulting 1D spectra after collapse. Both spikes are passed from the 2D to 1D spectra after collapse. (\textit{Bottom}) final median 1D spectrum containing the spike.}
    \label{fig:avg_12}
\end{figure}

Each source was observed three times as a result of the nodding pattern employed. We developed an averaging procedure that maximises the S/N of the spectra while also efficiently removing spikes and cosmic rays from the spectra that were missed during data reduction. Our averaging procedure works by comparing and averaging the 2D-rectified and 1D extracted spectra across the three nods. The 2D spectra are 2D arrays with seven pixels in the cross-dispersion direction (rows) and 3817 pixels in the dispersion direction (columns). Our procedure is as follows:
The 2D-rectified spectra are averaged together by calculating the median of each pixel in the 2D arrays. The resulting averaged 2D-rectified spectrum is then optimally extracted. We call this spectrum $S_1$. Next, the procedure goes back to the individual 2D-rectified spectra, and optimally extracts 1D spectra from them. These 1D spectra are then averaged into a single spectrum by finding the median of each wavelength bin. This spectrum we call $S_2$.\\ 
This produces two median spectra: $S_1$ and $S_2$. $S_1$ has typically lower S/N compared to $S_2$, but virtually all spikes are successfully removed during the averaging process. $S_2$ on the other hand shows typically higher S/N, but many spikes make it through the averaging process. The reason for $S_2$ being more resilient to spikes than $S_1$ is illustrated in figure \ref{fig:avg_12}. On the right of figure \ref{fig:avg_12}, which represents the averaging and collapsing process for $S_2$, we see that two of three 2D spectra contain a spike (red pixel). In this case, the spikes occur in the same detector column for each spectrum and hence at the same wavelength, but in different rows. When these 2D spectra are collapsed, the spikes pass through to the 1D spectrum. When these 1D spectra are averaged, since the spikes occur at the same wavelength, the final averaged spectrum also contains the spike. On the left of figure \ref{fig:avg_12}, representing the collapsing and averaging process for $S_1$, since the spikes are present in the same column, but not the same row, they are filtered out during the averaging process and hence are not passed to the collapsed 1D spectrum. For the spikes to pass through in this case, they would need to occur in both the same detector column and row, which is unlikely to occur since these spikes are random, due to cosmic rays and spurious pixels.

By comparing $S_1$ and $S_2$, the spikes in the higher S/N spectrum $S_2$ can be removed. This procedure was used for both the stellar spectra and the nebular spectra.
\newpage
\section{Fitted Hydrogen Lines} \label{app:fitted_lines}
\begin{figure*}[!hbt]
    \centering
    \onecolumn 
    \includegraphics[width=0.9\linewidth]{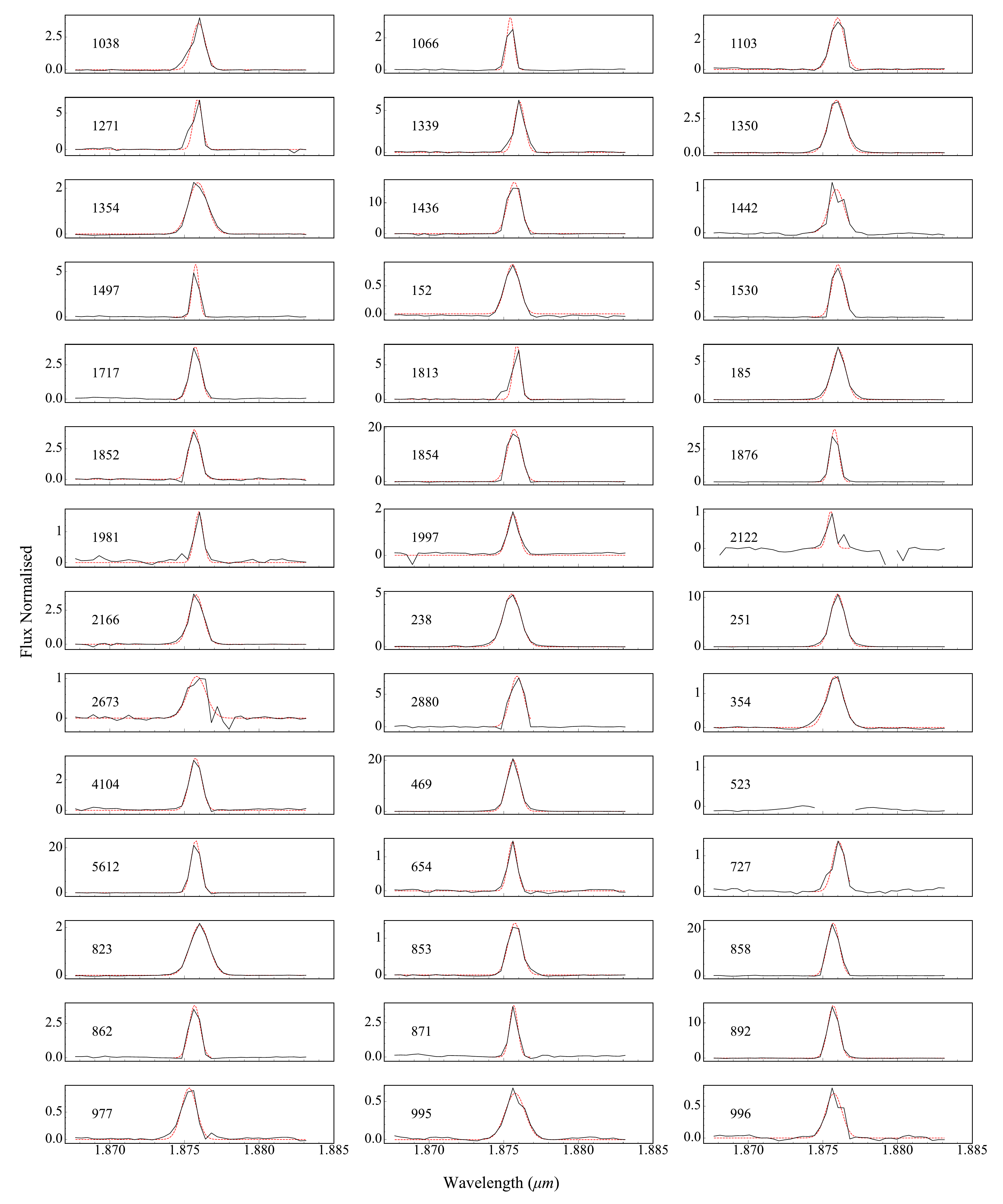}
    \caption{$Pa_{\alpha}$ emission line profiles of the PMS sources. The best fitting Gaussian profile is shown as a red line. Note that source 523 exhibits a central inversion in its emission line profile, and so no Gaussian fit was performed for this source.}
    \label{fig:Pa_a_grid}
\end{figure*}
\section{Emission line properties} \label{app:emission_line_properties}

\begin{table*}
\caption{\small{Emission line properties of the sample.}}
\small
    \centering
    \begin{tabular}{cccccc}
         \hline
         ID & EW $Pa_{\alpha}$ ($\AA$)& EW $Br_{\gamma}$ ($\AA$)&  EW $Br_{\beta}$ ($\AA$) & $\log(L_{acc})$ $(L_{\odot})$ & $\log(\dot{M}_{acc})$ $(L_{\odot})$\\
         \hline
         \hline
            727 & $-14.465 \pm 1.132$ & $0.077 \pm 0.013$ & $-8.756 \pm 0.688$ & $-1.634 \pm 0.327$ & $-8.986 \pm 0.149$ \\
            871 & $-26.073 \pm 2.092$ & $-2.092 \pm 0.169$ & $-7.254 \pm 0.54$ & $-1.162 \pm 0.671$ & $-8.537 \pm 0.34$ \\
            1497 & $-34.979 \pm 3.065$ & $-2.472 \pm 0.289$ & $-18.47 \pm 1.655$ & $-1.377 \pm 0.682$ & $-8.519 \pm 0.333$ \\
            2122 & $-7.811 \pm 0.624$ & $-1.861 \pm 0.171$ & $-3.095 \pm 0.266$ & $-1.176 \pm 0.671$ & $-8.35 \pm 0.297$ \\
            1813 & $-52.397 \pm 4.171$ & $-4.038 \pm 0.397$ & $-8.141 \pm 0.656$ & $-1.209 \pm 0.673$ & $-8.349 \pm 0.298$ \\
            654 & $-13.255 \pm 0.902$ & $-1.299 \pm 0.113$ & $-2.288 \pm 0.248$ & $-0.99 \pm 0.661$ & $-8.337 \pm 0.288$ \\
            2880 & $-89.532 \pm 6.832$ & $2.546 \pm 0.25$ & $-31.37 \pm 2.493$ & $-1.021 \pm 0.298$ & $-8.327 \pm 0.133$ \\
            1066 & $-18.534 \pm 1.396$ & $1.108 \pm 0.097$ & $-6.736 \pm 0.516$ & $-1.265 \pm 0.31$ & $-8.297 \pm 0.178$ \\
            1981 & $-12.408 \pm 1.258$ & $-4.912 \pm 1.541$ & $-7.099 \pm 1.016$ & $-1.271 \pm 0.677$ & $-8.251 \pm 0.295$ \\
            1852 & $-37.711 \pm 3.453$ & $-8.33 \pm 0.754$ & $-56.462 \pm 5.14$ & $-0.861 \pm 0.654$ & $-8.25 \pm 0.285$ \\
            1876 & $-275.524 \pm 22.986$ & $-8.488 \pm 0.662$ & $-107.521 \pm 7.992$ & $-0.94 \pm 0.659$ & $-8.204 \pm 0.335$ \\
            1436 & $-199.804 \pm 15.265$ & $-6.248 \pm 0.472$ & $-106.897 \pm 7.955$ & $-1.122 \pm 0.668$ & $-8.182 \pm 0.292$ \\
            1530 & $-80.695 \pm 5.858$ & $-3.305 \pm 0.275$ & $-35.209 \pm 2.758$ & $-1.057 \pm 0.665$ & $-8.171 \pm 0.36$ \\
            4104 & $-35.414 \pm 2.751$ & $-4.369 \pm 0.425$ & $-8.56 \pm 0.646$ & $-0.908 \pm 0.657$ & $-8.148 \pm 0.286$ \\
            1103 & $-40.155 \pm 3.015$ & $-5.497 \pm 0.459$ & $-21.01 \pm 1.568$ & $-0.958 \pm 0.66$ & $-8.103 \pm 0.298$ \\
            1717 & $-34.336 \pm 2.8$ & $-3.059 \pm 0.28$ & $-6.817 \pm 0.544$ & $-1.233 \pm 0.674$ & $-8.018 \pm 0.316$ \\
            996 & $-11.709 \pm 0.942$ & $-1.481 \pm 0.267$ & $-3.108 \pm 0.305$ & $-0.615 \pm 0.642$ & $-8.0 \pm 0.282$ \\\
            862 & $-34.583 \pm 2.421$ & $-2.554 \pm 0.195$ & $-16.981 \pm 1.366$ & $-0.916 \pm 0.657$ & $-7.964 \pm 0.308$ \\
            1271 & $-20.635 \pm 2.304$ & $-5.231 \pm 1.47$ & $-44.972 \pm 3.484$ & $-0.998 \pm 0.662$ & $-7.938 \pm 0.29$ \\
            1997 & $-16.628 \pm 1.329$ & $-2.605 \pm 0.445$ & $-3.123 \pm 0.337$ & $-0.639 \pm 0.643$ & $-7.907 \pm 0.326$ \\
            1339 & $-51.992 \pm 4.297$ & $-6.241 \pm 0.531$ & $-31.289 \pm 2.579$ & $-1.105 \pm 0.668$ & $-7.895 \pm 0.294$ \\
            2673 & $-18.046 \pm 1.378$ & $-3.286 \pm 0.432$ & $-10.029 \pm 0.735$ & $-0.651 \pm 0.644$ & $-7.893 \pm 0.281$ \\
            1854 & $-237.849 \pm 18.328$ & $-17.678 \pm 1.393$ & $-103.598 \pm 7.627$ & $-0.644 \pm 0.643$ & $-7.882 \pm 0.285$ \\
            1442 & $-13.45 \pm 1.08$ & $-1.36 \pm 0.198$ & $-4.556 \pm 0.494$ & $-0.788 \pm 0.651$ & $-7.844 \pm 0.288$ \\
            892 & $-147.971 \pm 10.994$ & $-13.017 \pm 0.98$ & $-57.864 \pm 4.223$ & $-0.32 \pm 0.627$ & $-7.687 \pm 0.287$ \\
            5612 & $-201.926 \pm 14.107$ & $-10.42 \pm 0.726$ & $-40.961 \pm 2.825$ & $-0.356 \pm 0.629$ & $-7.671 \pm 0.277$ \\
            523 & $-1.515 \pm 0.338$ & $0.201 \pm 0.039$ & $-0.207 \pm 0.173$ & $-0.803 \pm 0.288$ & $-7.492 \pm 0.135$ \\
            858 & $-204.169 \pm 16.219$ & $-21.418 \pm 1.929$ & $-96.943 \pm 10.664$ & $-0.065 \pm 0.39$ & $-7.425 \pm 0.182$ \\
            853 & $-17.685 \pm 1.245$ & $-2.226 \pm 0.213$ & $-4.967 \pm 0.356$ & $0.028 \pm 0.61$ & $-7.285 \pm 0.271$ \\
            2166 & $-45.029 \pm 3.29$ & $-7.817 \pm 0.62$ & $-10.372 \pm 0.753$ & $0.093 \pm 0.607$ & $-7.195 \pm 0.272$ \\
            977 & $-14.308 \pm 1.26$ & $-3.835 \pm 0.463$ & $-4.272 \pm 0.433$ & $0.184 \pm 0.603$ & $-7.126 \pm 0.264$ \\\
            1038 & $-42.392 \pm 3.797$ & $-3.423 \pm 1.463$ & $-5.47 \pm 1.036$ & $-0.37 \pm 0.629$ & $-7.056 \pm 0.279$ \\
            995 & $-12.522 \pm 1.306$ & $-2.71 \pm 0.596$ & $-3.521 \pm 0.411$ & $0.417 \pm 0.593$ & $-6.94 \pm 0.289$ \\
            354 & $-26.538 \pm 2.028$ & $-6.242 \pm 0.719$ & $-6.43 \pm 0.513$ & $0.795 \pm 0.578$ & $-6.53 \pm 0.254$ \\
            1354 & $-36.42 \pm 2.698$ & $-4.689 \pm 0.446$ & $-7.19 \pm 0.548$ & $0.755 \pm 0.579$ & $-6.424 \pm 0.255$ \\
            1350 & $-54.599 \pm 4.368$ & $-12.016 \pm 0.985$ & $-14.853 \pm 1.21$ & $0.635 \pm 0.352$ & $-6.42 \pm 0.192$ \\
            152 & $-12.989 \pm 0.958$ & $--$ & $-1.825 \pm 0.152$ & $0.861 \pm 0.22$ & $-6.255 \pm 0.101$ \\
            823 & $-36.491 \pm 3.227$ & $-6.914 \pm 0.616$ & $-8.896 \pm 0.789$ & $1.554 \pm 0.307$ & $-5.771 \pm 0.218$ \\
            469 & $-225.584 \pm 19.738$ & $-17.157 \pm 1.421$ & $-16.327 \pm 1.355$ & $2.074 \pm 0.536$ & $-5.256 \pm 0.235$ \\
            238 & $-74.174 \pm 6.14$ & $-16.326 \pm 1.491$ & $-23.958 \pm 2.007$ & $2.002 \pm 0.288$ & $-5.216 \pm 0.254$ \\
            251 & $-128.061 \pm 9.053$ & $-14.927 \pm 1.049$ & $-18.649 \pm 1.316$ & $2.3 \pm 0.53$ & $-4.943 \pm 0.301$ \\
            185 & $-78.229 \pm 6.861$ & $-10.586 \pm 0.915$ & $-13.754 \pm 1.199$ & $2.22 \pm 0.279$ & $-4.603 \pm 0.128$ \\

    \end{tabular}
    \tablefoot{\small{Sources are listed in terms of increasing $\dot{M}_{acc}$}. A number of sources shows net absorption line profiles for $Br_{7}$. In these cases we used equation \ref{revised_acc_lum} and $Pa_{\alpha}$ to determine $L_{acc}$ and $\dot{M}_{acc}$.}
    \label{tab:emission_line_properties}
\end{table*}

\end{appendix}

\end{document}